\documentclass[11pt]{article}
\usepackage[normalem]{ulem}
\pdfoutput=1 

\usepackage{jheppub} 
\usepackage{url}
\usepackage{caption}
\usepackage{subcaption}
\usepackage{extarrows}

\usepackage{hyperref}

\usepackage[latin9]{inputenc}
\usepackage{float}
\usepackage{amsmath}
\usepackage{amssymb}
\usepackage{graphicx}
\usepackage{esint}
\usepackage{hyperref}
\usepackage{comment}
\usepackage{color}
\usepackage{microtype}
\usepackage{cleveref}
\usepackage{breakurl}
\usepackage{bbm}
\newcommand{\be}{\begin{equation}}
\newcommand{\ee}{\end{equation}}
\newcommand{\ben}{\begin{displaymath}}
\newcommand{\een}{\end{displaymath}}
\newcommand{\bea}{\begin{eqnarray}}
\newcommand{\eea}{\end{eqnarray}}
\def\K{K{\"a}hler }
   \newcommand{\rf}[1]{(\ref{#1})}
\newcommand{\vp}{\varphi}

\def\be{\begin{equation}}
\def\ee{\end{equation}}
\def\bea{\begin{eqnarray}}
\def\eea{\end{eqnarray}}
\def\ba{\begin{array}}
\def\ea{\end{array}}
\def\bit{\begin{itemize}}
\def\eit{\end{itemize}}

\def\g{\gamma}

\def\d{\delta}

\def\a{\alpha}
\def\b{\beta}

\def\vp{\varphi}
\def\vt{\vartheta}

 \makeatletter

\allowdisplaybreaks

\makeatother

\makeatletter
\DeclareRobustCommand{\rcite}[1]{%
  \rcite@aux#1,\@nil{#1}%
}
\def\rcite@aux#1,#2\@nil#3{%
  \if\relax#2\relax
    Ref.~\cite{#3}%
  \else
    Refs.~\cite{#3}%
  \fi
}
\makeatother

\hypersetup{
    colorlinks = true,
    citecolor = {blue},
    linkcolor = {blue},
    urlcolor = {blue},
}

\usepackage{stackengine,amsmath}
\stackMath
\begin{document}

 \title{\rm {\bf \LARGE \boldmath  { Landscape of Modular Cosmology   }}}

\author{Renata Kallosh and }
\author{Andrei Linde}

\affiliation{Stanford Institute for Theoretical Physics and Department of Physics,\\ Stanford University, Stanford, CA 94305, USA}
\emailAdd{kallosh@stanford.edu}
\emailAdd{alinde@stanford.edu}

\abstract{We investigate the global structure of the recently discovered family of $SL(2,\mathbb{Z})$-invariant potentials describing inflationary $\alpha$-attractors. These potentials have an inflationary plateau consisting of the fundamental domain and its images fully covering the upper part of the Poincar\'e half-plane. Meanwhile, the lower part of the half-plane is covered by an infinitely large number of ridges, which, at first glance, are too sharp to support inflation. However, we show that this apparent sharpness is just an illusion created by hyperbolic geometry, and each of these ridges is physically equivalent to the inflationary plateau in the upper part of the Poincar\'e half-plane.
}

\maketitle


\section{Introduction}
\parskip 5pt

The idea that string theory inspired supergravity has a kinetic term with $SL(2, \mathbb{R})$ symmetry of the form
\be
{n\over 4} \, {\partial \tau \partial \bar \tau\over ({\rm Im}  \tau )^2}
\label{Ferrara} \ee
was proposed in \cite{Ferrara:1989bc}. The integer $n$ here is related to the \K curvature of $SL(2, \mathbb{R})/U(1)$ coset space\footnote{ Much later this same integer $3\alpha=n$ in case of cosmological $\alpha$-attractors was proposed and studied in \cite{Ferrara:2016fwe,Kallosh:2021vcf} as a target of the future cosmological experiments, like LiteBIRD  \cite{LiteBIRD:2022cnt}. These are 7 cases $n=1,2,3,4,5,6,7$ of Poincar\'e disks related to the upper half-plane by a Cayley transform.} which is $\mathbb{R}_K=-{2\over n}$. The proposal in \cite{Ferrara:1989bc} concerning supergravity potentials was that $SL(2, \mathbb{R})$ symmetry can be broken down, due to world-sheet instantons, to its discrete subgroup $SL(2,\mathbb{Z})$.
 This led to the conclusion that potentials depend on modular forms, like holomorphic modular function $j(\tau)$, holomorphic Dedekind function $\eta(\tau)$, and an almost holomorphic regularized Eisenstein function $\hat G_2 (\tau, {\rm Im} \tau)$. This symmetry of supergravity in \cite{Ferrara:1989bc} was dubbed ``target space modular invariance'' to distinguish it from duality symmetry in string theory.

$SL(2,\mathbb{Z})$ invariant cosmological theories are described by a four-dimensional bosonic actions
\be
S (\tau, \bar \tau) = \int d^4x  \sqrt{-g}\Big( {R\over 2} + {3\alpha\over 4} \, {\partial \tau \partial \bar \tau\over ({\rm Im}  \tau )^2}- V(\tau, \bar \tau)\Big)  \, .
\label{hyper}\ee
They depend on a single complex field $\tau(x) = \tau_1+i \tau_2$, where $\tau_1$ and $\tau_2$ are functions of  4 space-time coordinates $x^\mu$. Each term in the  action is  $SL(2,\mathbb{Z})$ invariant when
\be
\tau\to  {a \tau +b\over c \tau + d} , \qquad a,b,c,d \in \mathbb{Z}\, , \quad ad-bc=1 \ .
\ee 
It was proposed in \cite{Casas:2024jbw} to study $SL(2,\mathbb{Z})$ invariant cosmological theories with plateau potentials. Their potential was given by a specific function of the Dedekind function and $\tilde G_2$   Eisenstein modular form of weight 2.

In \cite{Kallosh:2024ymt}, we studied $SL(2,\mathbb{Z})$ cosmological plateau-type models of general type,  depending on $SL(2,\mathbb{Z})$ invariant Klein's $j$-function and  Dedekind function. We have shown that these models, as well as the model of Ref. \cite{Casas:2024jbw}, represent a novel class of $\alpha$-attractors studied earlier in \cite{Kallosh:2013yoa,Galante:2014ifa,Carrasco:2015uma,Carrasco:2015rva,Carrasco:2015pla,Kallosh:2021mnu}. 
We continue to study $SL(2,\mathbb{Z})$  inflationary models in 
 \cite{KalLin2024,CKLR}.
 
The early studies of target space modular invariance in \cite{Font:1990nt} revealed an important feature of modular invariant potentials: they have an infinity of degenerate minima,  saddle points, and maxima, whose positions are related to each other by modular $SL(2,\mathbb{Z})$ transformations. It was observed there that the choice of the minimum leads to a spontaneously broken target-space modular invariance. 

Cosmological inflation models based on $SL(2,\mathbb{Z})$ invariant action in Eq. \rf{hyper} were studied in \cite{Schimmrigk:2014ica,Schimmrigk:2016bde,Schimmrigk:2021tlv}. In particular, a model of $ j$-inflation was proposed in \cite{Schimmrigk:2014ica,Schimmrigk:2016bde,Schimmrigk:2021tlv}.  A global structure of the potential of $ j$-inflation was investigated in \cite{Schimmrigk:2021tlv}. It was observed there that the potential near the boundary at ${\rm Im} \, \tau \to 0$ has an intricate structure reminiscent of a fractal structure. But the contour plot of the potential has to be complimented by the behavior of the hyperbolic metric, which diverges as the saddles approach the real axis.

The difference between $ j$-inflation studied in  \cite{Schimmrigk:2014ica,Schimmrigk:2016bde,Schimmrigk:2021tlv} and the new $SL(2,\mathbb{Z})$ invariant models developed in \cite{Kallosh:2024ymt} is that $ j$-inflation  potentials at large $\tau_2$ grow exponentially in $\tau_2$, i.e. double exponentially in the canonically normalized inflaton field,  whereas in  \cite{Kallosh:2024ymt} the potentials depending on $j(\tau), \eta(\tau)$ approach a plateau.  Therefore, in $ j$-inflation,  inflation begins at a saddle point, whereas in  \cite{Kallosh:2024ymt}, inflation starts at the plateau, which has the same large-field behavior as the $\alpha$-attractor potentials.

\begin{figure}[H]
\centering
\includegraphics[scale=0.36]{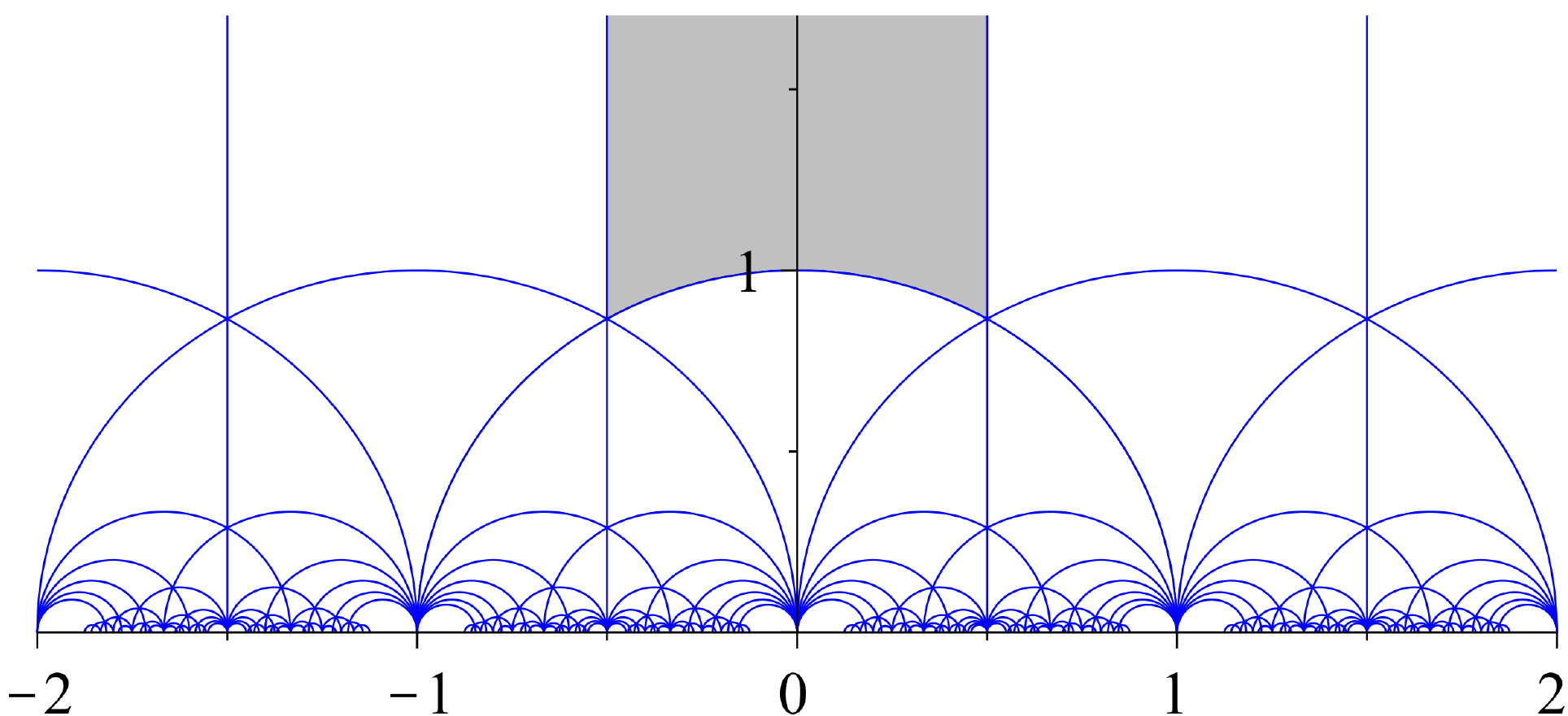}
\vskip -10pt
\caption{\footnotesize Tessellation of the hyperbolic  plane $\tau =x+iy$, $(\ -\infty <x < \infty, \, y>0)$ shows vertical bands which repeat each other. The grey area is a fundamental domain,  $-0.5 \leq x\leq 0.5$; $\tau \bar \tau \geq 1$, the other bands are shifted by 1 or -1 and repeated. There is a symmetry under reflection: $x\to -x$. The proliferation of the images of the points in the boundary of the fundamental domain in the grey area is a result of the eternal continuation of the geodesics, either vertical or semi-circular ones, towards the boundary $y\to 0$, see \href{https://en.wikipedia.org/wiki/Fundamental_domain}{Wikipedia}. }
\label{Tessellation}
\end{figure}

Here we will study the landscape of $SL(2,\mathbb{Z})$ cosmology.
The existence of this landscape, for example, of the infinite number of minima of the potentials,  or saddle points, is due to the fact that the value of the potential is the same after the modular transformation. Namely, any point $\tau'= {a\tau+b\over c\tau +d}$ with arbitrary numbers $a,b,c,d$ with $ad-cb=1$ is a  modular image of any original point $\tau$ since $V(\tau') = V(\tau)$. But the plateau of the potential is a big area; for example, in the fundamental domain in Fig.  \ref {Tessellation}, it is a grey area at large $y$.
From the first glance at the plot of the $SL(2, \mathbb{Z})$ 3D potentials in  \cite{Kallosh:2024ymt}, we find that in Cartesian coordinates, there is one plateau and many ridges.  This is a consequence of the fact that the metric in the hyperbolic space in Cartesian coordinates blows up near the boundary $y=0$ in the half-plane. Therefore, evaluating the shape of the potential without taking into account the metric may be misleading. In this paper, we will develop new tools which will help us to address this issue. 
Note that in the cosmological setting in space-time, as opposed to string theory on the world-sheet, $\tau= \theta+i \exp {\sqrt{2/3\alpha}\, \vp}$ is a space-time dependent complex scalar field $\tau(x^\mu)$. One should solve equations of motion, defining the evolution of scalars $\theta(t)$ and $\vp(t)$ in time. 

During our investigation of inflation starting at the inflationary plateau in the fundamental domain, we have found that after inflation, the scalar fields roll down and cross the boundary of the fundamental domain, its lower arc, where the minima of the potentials in \cite{Casas:2024jbw,Kallosh:2024ymt} are located. Their subsequent evolution, including the process of reheating, takes the scalars out of the fundamental domain. This is not surprising: the whole hyperbolic half-plane is geodesically complete, but the fundamental domain is not.

Moreover, as we have found out in \cite{Kallosh:2024ymt}, inflation in these models may begin outside of the fundamental domain. Thus, if we want to fully understand the cosmological evolution in this scenario, we should explore the global structure of the potential in the entire half-plane. This is the main goal of our investigation.

\section{Group of mappings}
In the context of $SL(2, \mathbb{Z})$ symmetry the Poincar\'e half-plain 
$
{\cal H}= \{ \tau \in \mathbb{C}  | \, {\rm \, Im} \tau  >0 \}
$
can be represented by a fundamental domain 
\be
I= \Big \{ \tau \in {\cal H} , |\tau| \geq 1, |{\rm \, Re} \tau | \leq {1\over 2} \Big \}
\ee
and its images. In our Figure \ref{Tessellation} the grey fundamental domain,  as well as its images, cover the whole Poincar\'e half-plain. One can see that these images arise via the map of the geodesics, which come closer and closer to the $y=0$ boundary, but never stop.  A short information on half-plane geodesics is given in Appendix \ref{Geodesics}.

The  algebraic nature of these images can be explained using the description in \cite{Rankin_1977} where the $SL(2, \mathbb{Z})$  matrices
\be
U= \left(\begin{array}{cc}1 & 1 \\0 & 1\end{array}\right)\, , \qquad V= \left(\begin{array}{cc}0 & -1 \\1 & \, 0\end{array}\right)\ ,
\ee
define the mappings  of any point $\tau$ as follows
\be
U\tau = \tau +1\, , \qquad V\tau = -1/\tau \ .
\ee
A matrix $P$ is now introduced\footnote{Often a different notation is used, $U=T, V=S$ and therefore $P= ST$.},  and $P^2$ and $P^6$ follow
\be
P= VU =\left(\begin{array}{cc}0 & -1 \\1 & \, 1\end{array}\right) \, , \qquad P^2= \left(\begin{array}{cc}-1 & -1 \\1 & \, 0\end{array}\right) \, , \qquad P^6 =I \ .
\ee
$SL(2, \mathbb{Z})$ mapping is generated by $V$ and $P^2$ in a unique way: the action of the operators $T[r, p_0, \dots, p_n]$ on any point $\tau$ will bring all of its modular images
\be
T[r, p_0, \dots, p_n] \, \tau = (-1)^r P^{2p_0} V P^{2p_1} V\dots V P^{2p_n} \, \tau  \ ,
\label{un}\ee
where 
\be
0\leq r\leq 1\, , \quad 0\leq p_i\leq 2 \, , \quad 0\leq i \leq n \, , \quad p_i>0\, , \quad 0 <  i < n \ .
\ee
Clearly, there is an infinite amount of images for any initial value of $\tau$ since the integers $n$ grow unrestricted. Many other mappings are known but they are typically not unique, as different from Eq. \rf{un}. For example, one can use 
\be
T[q_0; q_1, \dots, q_n]= U^{q_0} V U^{q_1} V\dots V U^{q_n} \, , \qquad q_i \in \mathbb{Z}\, , \qquad 0\leq i \leq n  \ .
\label{nonun}\ee
It is useful to know that all modular images are defined in Eq. \rf{un} in a unique way.  
In what follows, however, for practical purposes of showing examples of the first few steps of defining the images of the saddle point, we will use the set of operators in Eq. \rf{nonun}. All operators in Eq. \rf{nonun} are known to represent every element of SL(2,Z), although it is not a unique representation, but includes all of them, according to \cite{Rankin_1977}.

\begin{figure}[H]
\centering
\includegraphics[scale=0.18]{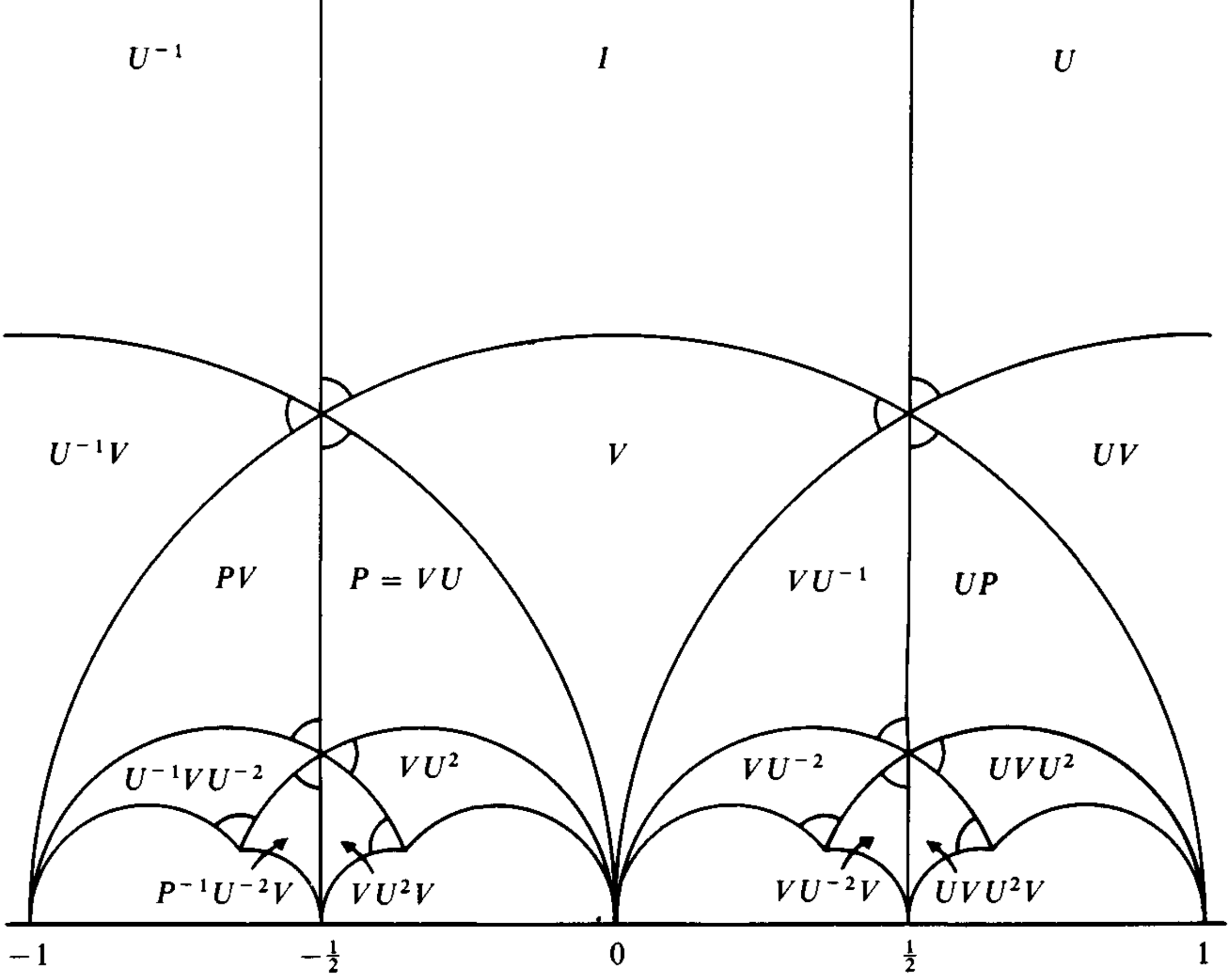}
\vskip -10pt
\caption{\footnotesize The standard fundamental domain I for the $SL(2,\mathbb{Z})$ in half-plane $\cal{H}$ and its images realized by the action of various $SL(2,\mathbb{Z})$ operations $V, U, P$ as given in \cite{Rankin_1977}. For example, the image of the fundamental domain after one inversion $V$ becomes highly distorted: two parallel boundaries of I going to $y\to \infty$  after inversion transform into two arcs merging in the limit $y\to 0$.
}
\label{Rankin}
\end{figure}
In Fig. \ref{Rankin}, we show how various $SL(2,\mathbb{Z})$ transformations generate the images of the fundamental domain I.
We will find out later that the set of operations in Eq. \rf{nonun} is particularly useful to define a proliferation of saddle points and show that it is related to continued fractions. This, in turn, will allow us to find a new set of Killing coordinate systems, which will allow us to establish equivalence between the sharp ridges and large inflationary plateaus.

\section{Cartesian and Killing coordinates on hyperbolic half-plane} \label{Sec:cartesian}
\subsection{Translation and reflection manifest actions in Cartesian coordinates }
Consider  Cartesian coordinates on the half-plane
\be
\tau = x+iy \, ,\qquad y >0 \ .
\ee
Kinetic term in these coordinates is\footnote{Our notation means that $(\partial  x)^2$ is for $(\dot x)^2 -(x')^2$.}
\be
{3\alpha\over 4} \, {\partial \tau \partial \bar \tau\over ({\rm Im}  \tau )^2}= {3\alpha\over 4} \, { (\partial  x)^2+ (\partial  y)^2\over y^2 } \ .
\label{kin}\ee

 All isometries of this hyperbolic geometry were listed in \cite{Carrasco:2015uma}. These include the $SL(2,\mathbb{Z})$ translation
\be
\tau \to \tau + 1 \, : \qquad x\to x+1\, ,  \qquad y\to y\, ,
\label{transl}\ee
and inversion
\be
\tau \to -{1\over \tau}  \, : \qquad x+iy \to -{1\over x+iy}\, .
\ee
There is also a reflection symmetry,  but only for $x$ since $y>0$, 
\be
\tau \to - \bar \tau\, , \qquad x\to -x,\, ,  \qquad y\to y\, .
\label{ref}\ee
Thus, kinetic term in Cartesian coordinates $x={\rm Re} \, \tau, y=  {\rm Im} \, \tau$ is manifestly symmetric under translation and reflection of ${\rm Re} \, \tau$.

Fig. \ref{Tessellation} shows the tessellation of the hyperbolic plane in $x, y$ coordinates.
Various vertical bands differ by translation given in Eq. \rf{transl}
and one can also see a reflection symmetry given in Eq. \rf{ref}.

For $SL(2,\mathbb{Z})$  invariant potentials depending on $|j(\tau)|^2$ or $|\eta(\tau)|^2$ we can see that both potentials have a reflection symmetry \rf{ref}. To see it, we remind that 
\be
j(\tau) \to j(q= e^{2\pi i\tau})\ , \qquad \overline {j(\tau)} = j(\bar q=e^{-2\pi i \bar \tau })  \ .
\ee
This means that 
\be
j(\tau) \overline {j(\tau)}  =  j(q= e^{2\pi i\tau})\, j(\bar q=e^{-2\pi i \bar \tau })  \ .
\ee
Thus, the potentials which depend on $j(\tau) \overline {j(\tau)}  $ have a reflection symmetry under
\be
\tau\to -\bar \tau
\ee
According to Eq. \rf{ref}, this means that 
\be
V(x, y) = V(-x, y) \ .
\ee
We can see the confirmation of this symmetry in all plots of $SL(2,\mathbb{Z})$ potentials  in \cite{Kallosh:2024ymt} and in Figs. \ref{Cartesian} - \ref{3Ridges} here.

We can define 
\be
y=e^{\sqrt{2\over 3\alpha}  \vp } >0\, , \qquad x\equiv \theta \ .
\ee
The horizontal line 
\be
y=1\, ,  \qquad \vp=0
\ee
splits the half-plane into upper part, which is unrestricted with   $y\to \infty$,
\be
\vp\to \infty \, , \qquad y>1  \qquad \vp>0\ ,
\ee
and lower part, which can reach the lower  boundary $y\to 0$ with 
\be
\vp\to -\infty\, ,  \quad 
 y<1  \qquad \vp<0 \ .
\ee
At $y=1$ both $\sqrt{3\a\over 2} x$ and $\sqrt{3\a\over 2} y$ are canonical as one can see in Eq. \rf{kin}. When we will plot our 3D potentials in $\theta, \vp$ coordinates, we will see manifest translation symmetry in $\theta$ 
\be
V\Big (\theta, \, e^{\sqrt{2\over 3\alpha}  \vp }\Big ) = V\Big (\theta+n, \, e^{\sqrt{2\over 3\alpha}  \vp }\Big ) \ ,
\ee
and kinetic term is also translationally invariant.

\be
{3\alpha\over 4} \, {\partial \tau \partial \bar \tau\over ({\rm Im}  \tau )^2}= {1\over 2} \,  \Big ( (\partial  \vp)^2 +  {3\alpha\over 2} e^{-2\sqrt{2\over 3\alpha}\vp}(\partial  \theta)^2 \Big ) \ .
\label{kinaxin}\ee
However, potentials at positive and negative $\vp$ are very different. Inversion symmetry is not manifest, namely, 
\be
\tau \to -{1\over \tau} \, : \qquad x+iy \to - {1\over x+iy} \ .
\ee
For example, at $x=0$ it means $y\to y^{-1}$ and $\vp \to -\vp$ but at $x\neq 0$ inversion  is complicated.
Translation and inversion symmetry together define modular images of the points inside the fundamental domain and at its boundaries to outside, as shown in Figs. \ref{Tessellation}, \ref{Rankin}.

\subsection{Inversion and reflection  manifest actions in Killing coordinates }\label{Sec:Kil}
We define  \be
\tau = i e^{\Phi} \, , \quad \Phi\in \mathbb{C} \ .
\label{original}\ee
Consider polar-type coordinates   $\Phi= \sqrt{2\over 3\alpha} (\tilde \vp -i \vt)$ in the half-plane, in the form
\be
\tau = i e^{\sqrt{2\over 3\alpha} (\tilde\vp -i\vt)} = e^{\sqrt{2\over 3\alpha} \tilde\vp}\Big (\sin (\sqrt{2\over 3\alpha} \vt)  +  i \cos (\sqrt{2\over 3\alpha} \vt )\Big ) \ ,
\label{Kil}\ee
and 
\be
-\infty <\vp < \infty \, , \qquad   -\pi/2 < \sqrt{2\over 3\alpha} \vt < \pi/2 \ .
\ee
These coordinates are not really polar since the angle $\vt$ is constrained to keep $y$ positive.
The kinetic term in these coordinates is
\be
{3\alpha\over 4} \, {\partial \tau \partial \bar \tau\over ({\rm Im}  \tau )^2}= {1\over 2}{ (\partial \tilde\vp)^2+ (\partial\vt)^2\over \cos^2 (\sqrt{2\over 3\alpha} \vt)} \ ,
\label{kinK}\ee
to be contrasted with the one in \rf{kinaxin}.
We have introduced these coordinates in  \cite{Carrasco:2015rva,Carrasco:2015pla},  where we called them Killing adapted coordinates since the metric does not depend on the inflaton $\tilde\vp$. At $\vt=0$ both fields $\tilde\vp, \vt$ are  canonical. 

 The relation between $\vp, \theta$ coordinates and $\tilde\vp, \vt$ is the following
  \be\label{vp}
y=e^{\sqrt{2\over 3\alpha}  \vp }=   e^{\sqrt{2\over 3\alpha} \tilde\vp} \cos \Big (\sqrt{2\over 3\alpha} \vt \Big) \ ,
   \ee
   \be
x=\theta=    e^{\sqrt{2\over 3\alpha} \tilde\vp} \sin \Big (\sqrt{2\over 3\alpha} \vt \Big) \ ,
 \label{theta} \ee
or
\be
{\tau_1\over \tau_2} = \tan  \Big (\sqrt{2\over 3\alpha} \vt \Big)\, , \qquad \tau \bar \tau= e^{2 \sqrt{2\over 3\alpha}\tilde \vp} \ .
\ee  
Note that $\tilde\vp = \vp$ for $\vt = 0$, and $ \theta=\sin  \Big (\sqrt{2\over 3\alpha} \vt \Big)$ for  $\tilde \vp= 0$.

The symmetry under inversion is
\be
\tau \to -{1\over \tau} \, : \qquad \Phi \to -\Phi \qquad \rightarrow  \qquad \, \tilde \vp\to -\tilde \vp\, , \quad  \vt\to -\vt\, \ .
\label{inv}\ee
When we will plot our 3D potentials in $\tilde \vp, \vt$ coordinates, we will see manifest symmetry with respect to the transformation 
$\tilde  \vp\to -\tilde \vp\, ,   \vt\to -\vt
$.
Thus, the potentials in these coordinates will be manifestly inversion invariant.
  The symmetry under translation in Killing coordinates will not be manifest.
  
Inversion symmetries, a priory, were expected to be only at simultaneous action on both fields in Eq. \rf{inv}. However, in Cartesian coordinates, we have shown that the theory is also invariant under reflection symmetry, $\theta\to -\theta$, see Sec. \ref{Sec:cartesian}. This translates into a symmetry 
$
\vt \to -\vt
$,
according to Eq. \rf{theta}. Therefore, we have independent symmetries in Killing coordinates, 
\be\tilde \vp\to -\tilde \vp  \quad \rm{and}, \quad \rm{separately}, \quad   \vt \to -\vt
\ee

\section{$SL(2, \mathbb{Z})$ invariant potentials}
\subsection{Cartesian coordinates}\label{cart}
\begin{figure}[H]
\centering
\includegraphics[scale=0.6]{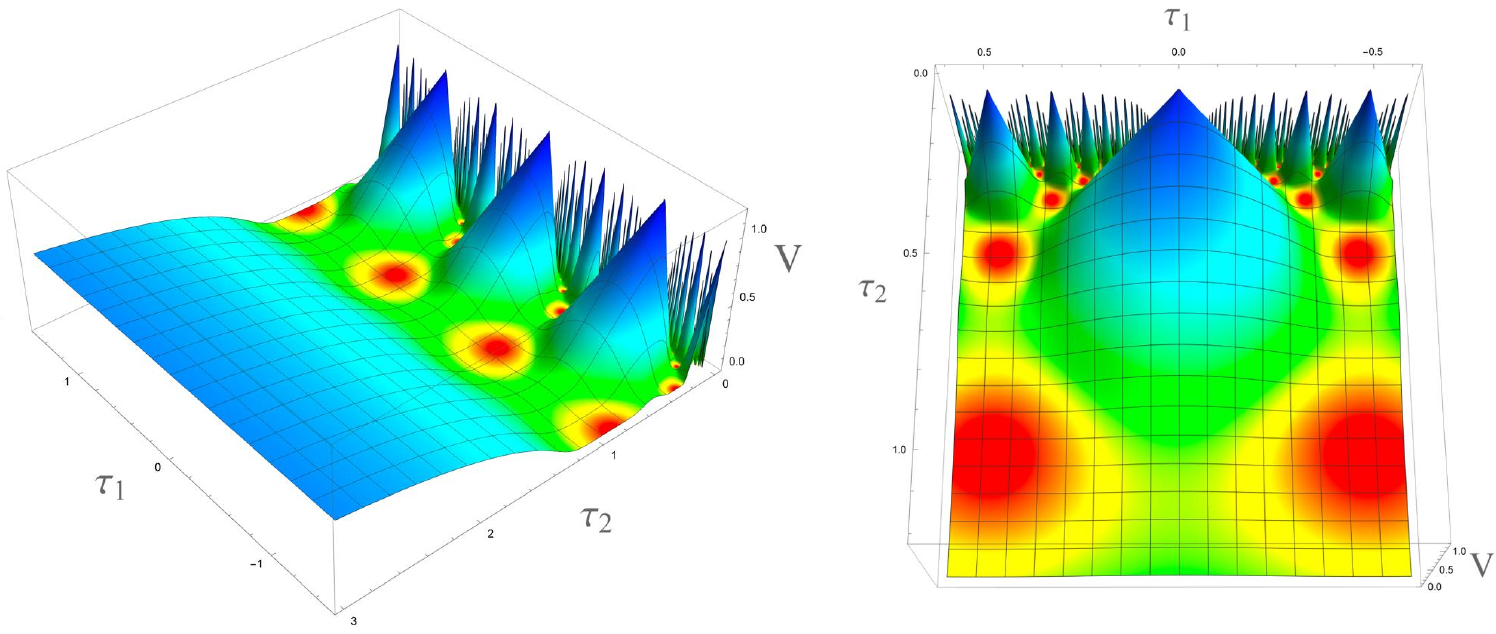}
\vskip -10pt
\caption{\footnotesize  Potential \rf{Renata2} as a function of $\tau_1$ and $\tau_2$. The blue plateau at the left panel's left side is where inflation may begin.  In the left panel, we show the potential for $-0.6<\tau_{1}< 0.6$, $\tau_{2}< 1.2$. It looks like a green field with pyramids growing there. The minima of the potentials are shown by red, and the saddle points are shown by light yelow-green paths between different minima.}
\label{Cartesian}
\end{figure}

Consider one of the potentials studied in  \cite{Kallosh:2024ymt}
\be\label{Renata2}
V(\tau, \bar \tau)= V_0 \Big (1-{ \log |j^2(i)|
\over  \ln ( | j(\tau))|^2 +  j^{2}(i))}\Big) \ .
\ee
 We plot this potential as a function of $\tau_1$ and $\tau_2$, where  $\tau =\tau_1+i\tau_2$, for $V_{0}= 1$, see Fig.  \ref{Cartesian}.
 
 The long blue plateau at the left panel of Fig.  \ref{Cartesian} is a  part of an infinite inflationary plateau at $\tau_2 \gg 1$. It plays the main role in the investigation of inflation in this model \cite{Kallosh:2024ymt,KalLin2024}, where the structure of this plateau was examined in detail. 
 
In this paper, we will concentrate on understanding the structure of the rest of this figure, which looks like a forest of pyramids growing increasingly narrow and sharp at $\tau_{2} \to 0$. The sharpness is due to the singularity of metric ${(\partial \tau_1)^2\over (\tau_2)^2}$ at $\tau_{2} \to 0$. These pyramids are surrounded by a series of red spots corresponding to a large number of minima of the potential of equal depth $V = 0$; see the right panel of Fig. \ref{Cartesian}. 

\begin{figure}[H]
\centering
\includegraphics[scale=0.35]{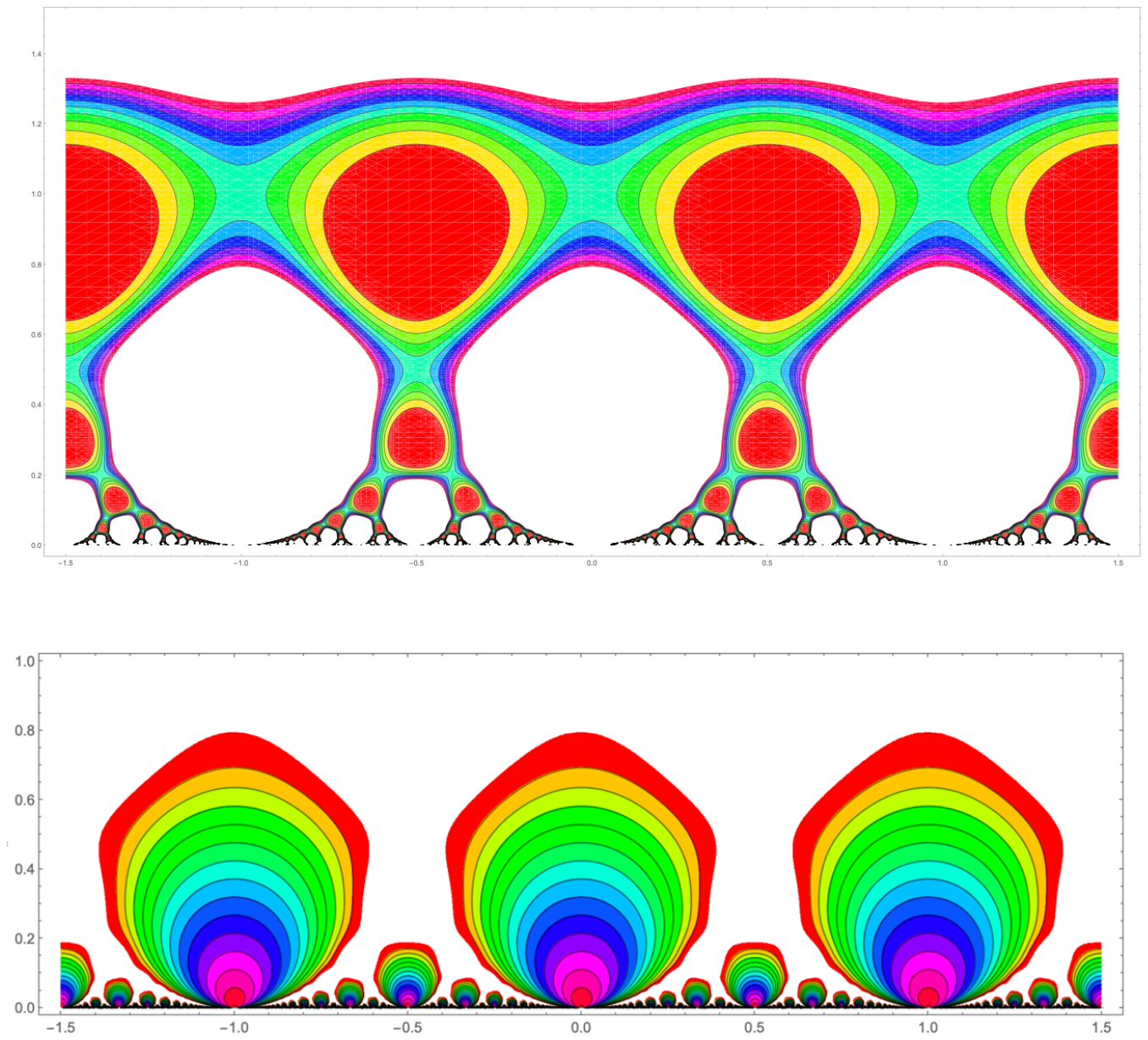}
\vskip -5pt
\caption{\footnotesize  The contour plot of the  potential \rf{Renata2} as a function of $\tau_1$ and $\tau_2$. The upper panel shows areas where  $0<V<0.1$. The red circles show the regions close to the minima with $V= 0$. The lower panel shows the higher parts of the potential, with $0.1<V<1$. The maxima of the potential are indicated by tiny red disks at the top of the potential with $V = 1$ at the boundary $\tau_{2} \to 0$.}
\label{3domains}
\end{figure}

 Most of the red spots (minima) are hidden by the pyramids, and some of the pyramids are hidden by others. To reveal them, we show two contour plots. The upper panel of Fig. \ref{3domains}  shows only the lower part of the potential, with $0<V<0.1$; the red spots correspond to the vicinities of the minima with the potential vanishing at the bottom of each minimum.
The lower panel of Fig. \ref{3domains}  shows only the higher part of the potential, at $\tau_2 < 1$,  $0.1<V<1$. They represent the pyramids, each of which has a maximum shown by tiny red disks at $V = 1$, at the boundary of the moduli space $\tau_2 \to 0$.  It is apparent from these figures that the potential forms an infinite fractal landscape. Naively, one would expect that only the non-fractal part of the potential, the part with $\tau_{2} \gg 1$, is suitable for inflation. However, as we will see, this is an illusion: In proper coordinates, each of the peaks in this fractal landscape represents an inflationary plateau.

This is a very counter-intuitive statement, so we will need to make several steps using some other coordinate systems, which may help us understand what is happening.

\subsection{Axion-inflaton coordinates}
As a first step, we make a change of variables $\tau_{1} = \theta$, $\tau_{2} = e^{\sqrt{2\over 3\alpha}  \vp }$, where $\theta = \tau_{1}$ is the axion field, and $\vp$  is the canonically normalized inflaton field. In these coordinates, $\theta$ is canonically normalized only at $\vp = 0$.

\begin{figure}[h!]
\centering
\includegraphics[scale=0.17]{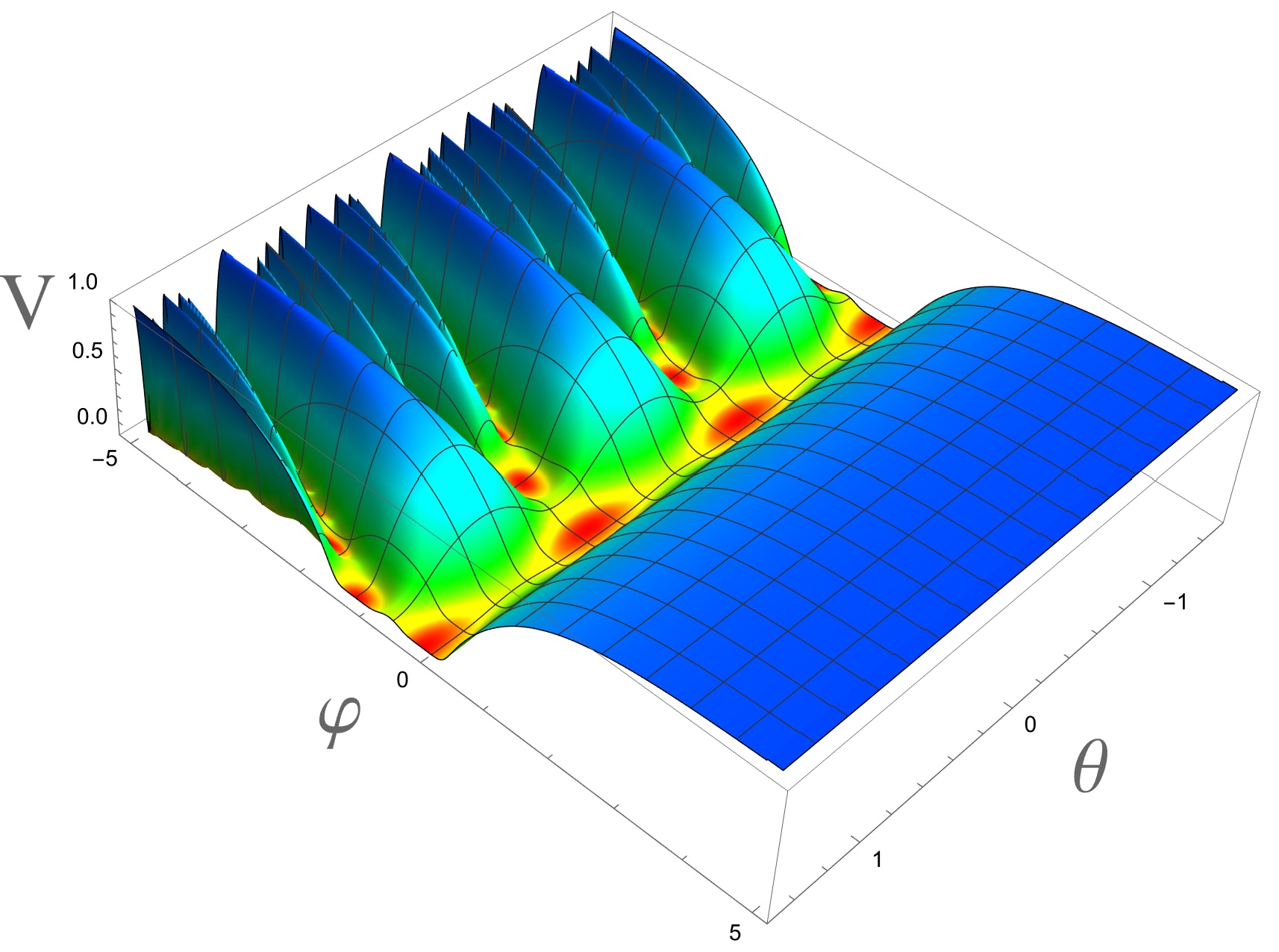}
\vskip -10pt
\caption{\footnotesize  Our potential \rf{Renata2} depending on $\tau = \theta+ i e^{\sqrt{2\over 3\alpha}  \vp }$. For comparison with the left panel of Fig. \ref{Cartesian}, we show the potential in 3 bands,  $-1.5 <\theta <1.5$ for $V_{0} = 1$  and $3\alpha=2$. One can clearly see the plateau at positive $\vp$, the minima at $\theta = -1.5,  -0.5,  0.5,  1.5$, and   saddle points at $\theta=-1,   0,   1$. At  $\vp < 0$, one can see a complicated profile of multiple mountains and a proliferation of minima and saddle points.}
\label{3band}
\end{figure}

The potential in these coordinates is shown in  Fig. \ref{3band}.

This coordinate system is quite suitable for the investigation of inflation in this model. First of all,  the potential is periodic with period $\Delta \theta = 1$. Therefore it is sufficient to study initial conditions in the fundamental domain with $-0.5\leq\theta  \leq 0.5$. Secondly, during inflation at large positive $\vp$, the field $\theta$ practically does not change.  

Indeed, according to \cite{Kallosh:2024ymt,KalLin2024}  the   derivatives of the potential  \rf{Renata2} with respect to the inflaton field $\vp$ are suppressed by the exponential factor $e^{-\sqrt{\frac{2}{3\alpha}}\vp}$, whereas the derivatives with respect to the axion field $\theta$ are suppressed much stronger, by a {\it double-exponential}  factor  $e^{-2\pi\, e^{\sqrt{\frac{2}{3\alpha}}\vp}}$.
 Therefore the field $\theta$ practically does not move during inflation at $\vp \gg \sqrt \alpha$, and inflationary predictions of this class of theories are   similar to the predictions of the single-field $\alpha$-attractors  \cite{Kallosh:2024ymt,KalLin2024,CKLR}. 

But these simplifications at $\vp \gg \sqrt \alpha$ do not simplify the structure of the potential at $\vp \ll 0$ shown in Fig. \ref{3band}. It reveals new and new ridges when we look at the potential at large negative values of $\vp$. Instead of looking like pyramids as in Figs. \ref{Cartesian}, these ridges look more like shark fins. 

\begin{figure}[H]
\centering
\includegraphics[scale=0.16]{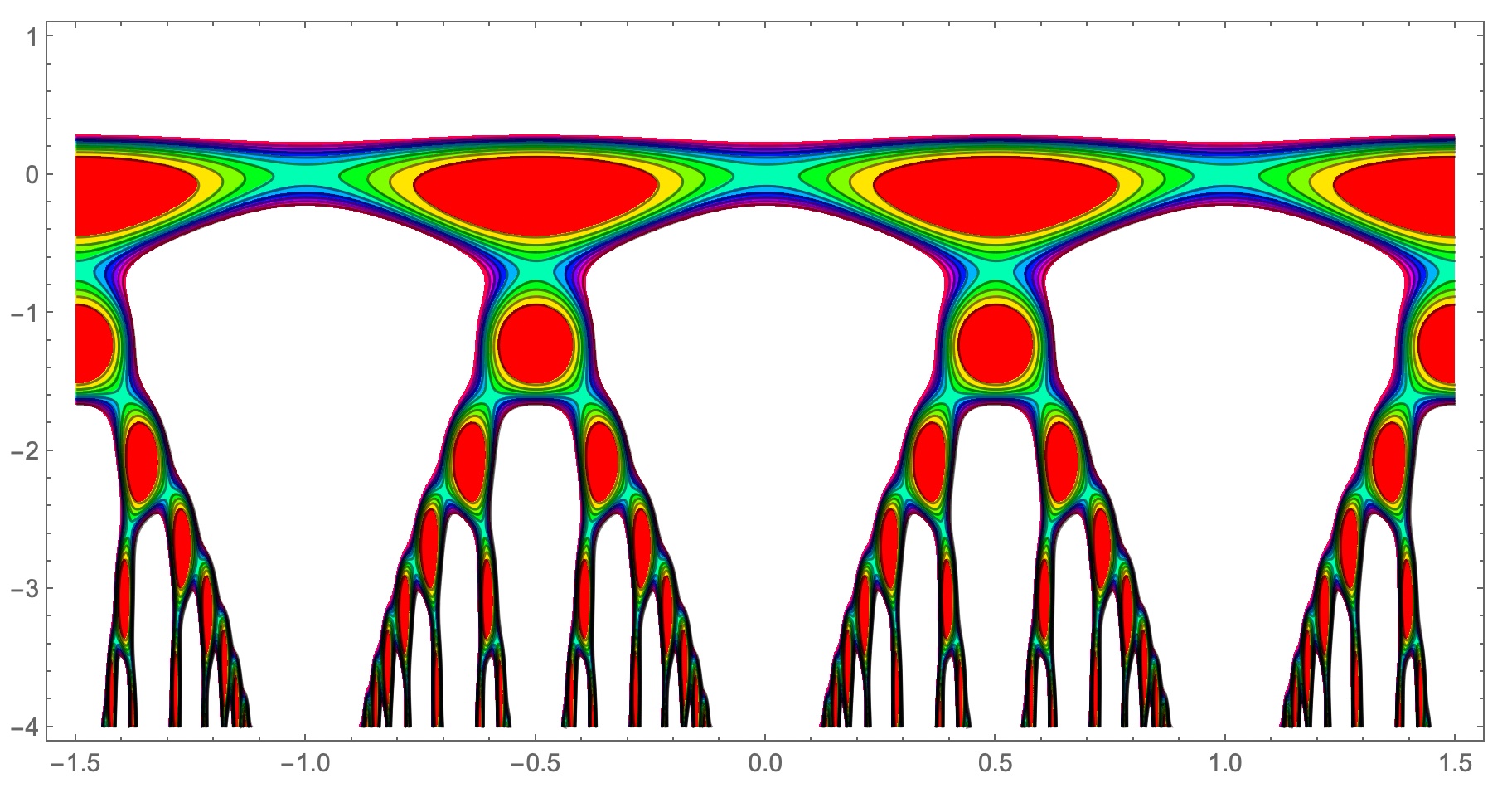}
\vskip -10pt
\caption{\footnotesize  The contour plot of the  potential \rf{Renata2} as a function of $\theta$ and $\vp$, for $-1.5<\theta<1.5$ and $-4<\vp< 1$. This figure, just as Fig. \ref{3domains}, shows areas where  $0<V<0.1$. The red ellipses show the regions close to the minima with $V= 0$.}
\label{3Ridges}
\end{figure}

As before, the ridges in Fig. \ref{3band} hide some of the minima of the potential shown by red. Their fractal distribution is shown in the contour plot Fig. \ref{3Ridges}. The main difference with the distribution in Cartesian coordinates shown in Fig. \ref{3domains} is that previously the lower region with $\tau_{2} < 1$ had a finite width $0<\tau_{2}<1$, whereas now it is stretched all the way down to $\vp \to -\infty$. This results in the stretching down of the fractal landscape in Fig. \ref{3Ridges}.

\begin{figure}[h!]
\centering
\includegraphics[scale=0.17]{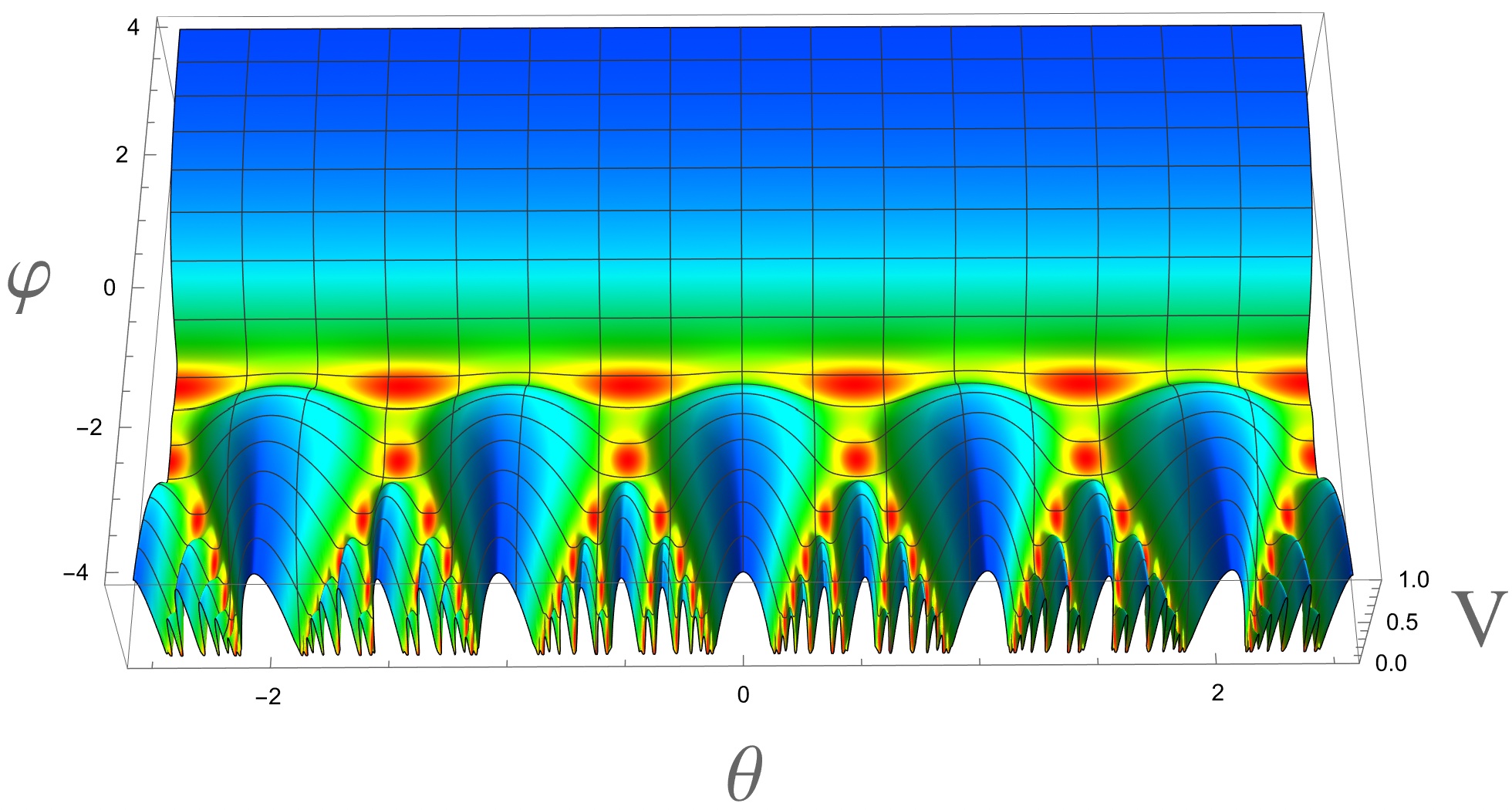}
\vskip -10pt
\caption{\footnotesize  Our potential \rf{Renata2} depending on $\tau = \theta+ i e^{\sqrt{2\over 3\alpha}  \vp }$ of the kind we show in Fig. \ref{3band}. Here we show five bands with saddle points at  $y=i$, $\theta=n$ bridging the plateau with five ridges. The nearest to plateau red minima are at $\tau_2= \sqrt{3/4}, \tau_1= 1/2+n$}
\label{5band}
\end{figure}

In the previous figures, we plotted the potential in a large range of $\theta$ to illustrate the periodicity of the potential. Note that there are saddle points of the potential at $\tau = n+i$, see Figs. \ref{Cartesian}-\ref{3Ridges}. They are positioned at the centers of ``bridges'' connecting the inflationary plateau with $\vp > 0$ and pyramids or ridges starting to grow from the points $\tau = n+i$  in the direction of large negative $\vp$.  See also Fig. \ref{5band} with five bands.

\begin{figure}[h!]
\centering
\includegraphics[scale=0.15]{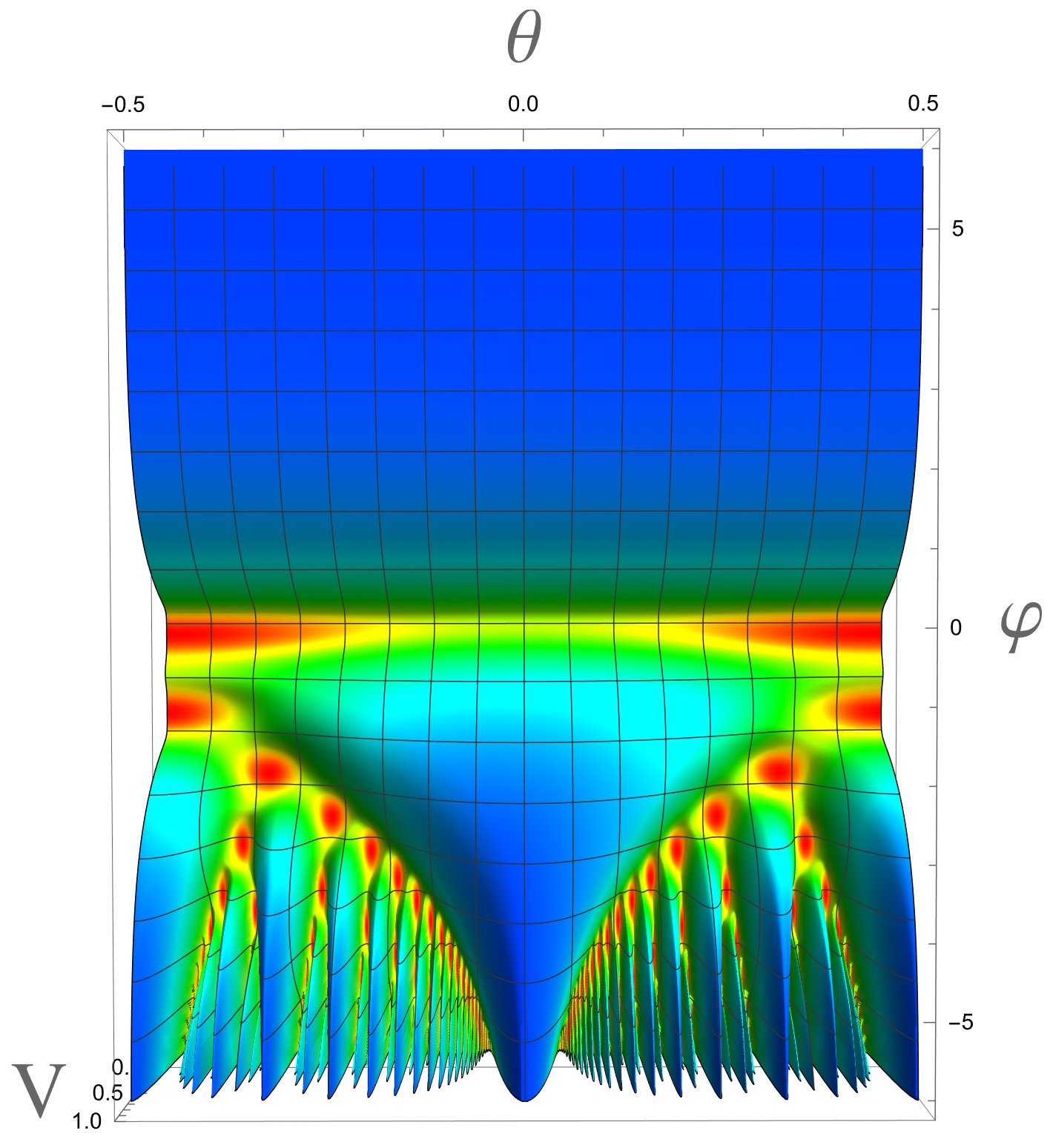}
\vskip -10pt
\caption{\footnotesize  Top view on the potential \rf{Renata2} as a function of $\theta$ and $\vp$ for $-0.5< \theta  < 0.5$, $V_{0} = 1$  and $3\alpha=2$. There is an inflationary plateau at  $\vp > 0$, the minima at $\theta  =  -0.5,  0.5$, and a saddle point at $\theta = 0$. At  $\vp < 0$, one can see a complicated profile of multiple mountains and a proliferation of minima and saddle points.}
\label{1Ridge}
\end{figure}

In the next figure, we simplify the figure slightly and focus on the fundamental domain with  $-0.5\leq\theta  \leq 0.5$ in Fig. \ref{1Ridge}.


None of the figures shown above suggests a possibility of inflation during rolling from the sharp top of a pyramid or by riding one of the shark fins. And yet, this possibility is there, but it is well hidden because of the coordinate systems' peculiarities.  We are going to address this problem in the next subsection.

\subsection{Killing coordinates}
Now, following \cite{Kallosh:2024ymt},  we will study the same potential \rf{Renata2} as in Fig. 
\ref{1Ridge}, but in Killing coordinates 
\be
\tau = i e^{\sqrt{2\over 3\alpha} (\tilde\vp -i\vt)} \ ,
\label{Killing}\ee
 where
 the kinetic term  is
$ {1\over 2}{ (\partial \tilde\vp)^2+ (\partial\vt)^2\over \cos^2 (\sqrt{2\over 3\alpha} \vt)} 
$, see Sec. \ref{Sec:Kil} for details. Let us find out what happens with regions $y\to \infty$ and the boundary at $y=0$ in Killing coordinates.

At  large $y$  in Cartesian and Killing coordinates, we find that
\be
y \to \infty :  \qquad e^{\sqrt{2\over 3\alpha} \tilde\vp} \cos \Big (\sqrt{2\over 3\alpha} \vt \Big)\to \infty \quad \Rightarrow \quad \tilde\vp \to \infty \ .
\ee
At the boundary $y\to 0$ in Cartesian and Killing coordinates at $|\sqrt{2\over 3\alpha} \vt | <  {\pi\over 2}$ we find that 
\be
y \to 0 :  \qquad e^{\sqrt{2\over 3\alpha} \tilde\vp} \cos \Big (\sqrt{2\over 3\alpha} \vt \Big)\to 0 \quad \Rightarrow \quad \tilde\vp \to - \infty \ ,
\ee
but at any particular finite $\tilde\vp$ we have
\be
y \to 0 :  \qquad e^{\sqrt{2\over 3\alpha} \tilde\vp} \cos \Big (\sqrt{2\over 3\alpha} \vt \Big)\to 0 \quad \Rightarrow \quad \vt \to \pm {\pi\over 2}  \ .
\ee
This means that in Killing  coordinates the Cartesian boundary $y \to 0$ is either at $\tilde\vp \to - \infty$ or, for any finite $\tilde \vp$, at $\vt \to \pm {\pi\over 2} $.
\begin{figure}[H]
\centering
\includegraphics[scale=0.45
]{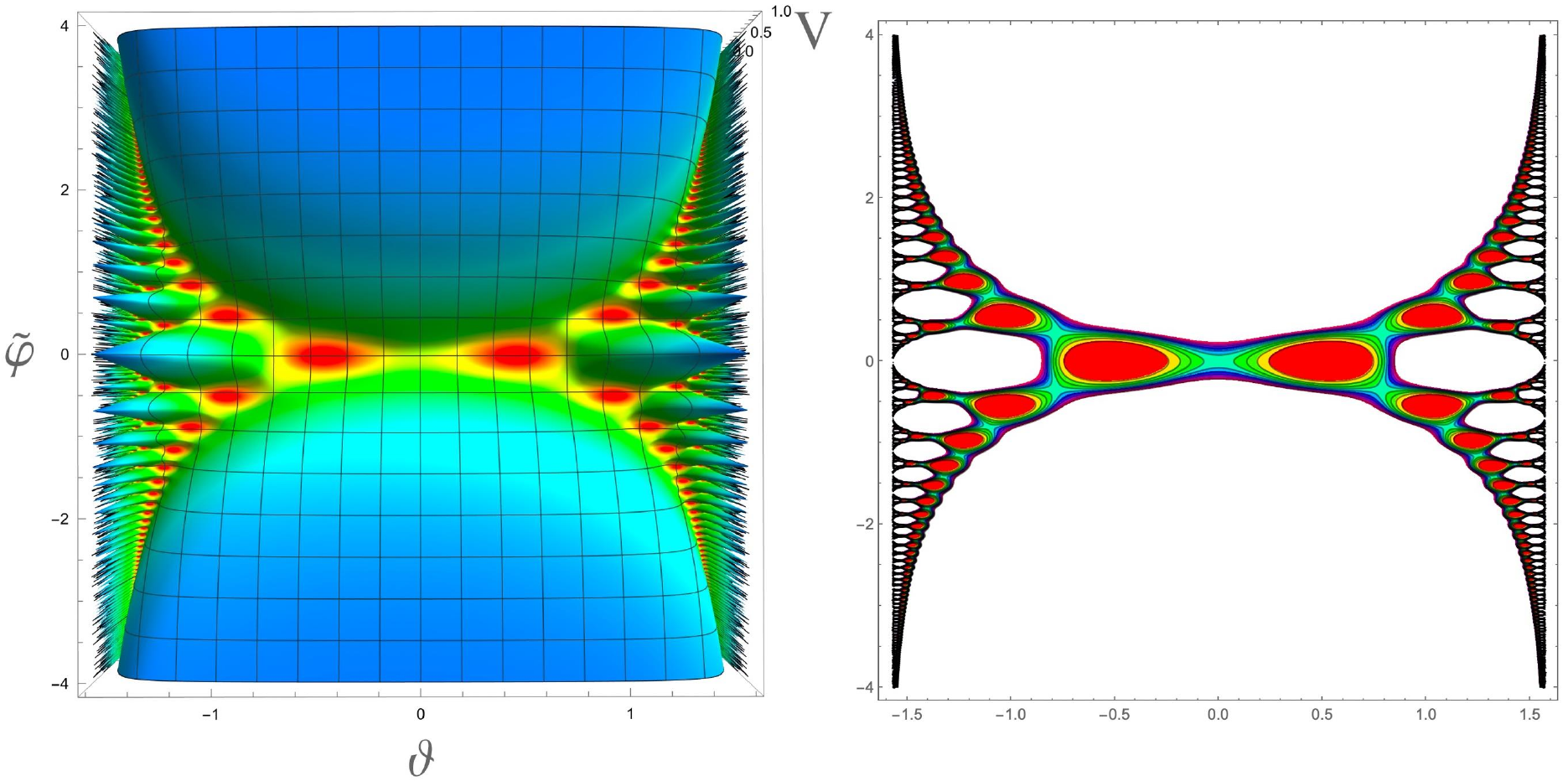}
\caption{\footnotesize Potential \rf{Renata2} for $V_{0} = 1$,   $3\alpha=2$,  in the coordinates \rf{Killing}.  In the left panel, we show the top view of the 3D plot of the potential. In the right panel, we show the contour plot of this potential in the range $0<V< 0.1$. This allows to see the full chain of red minima and yellow-green saddle points separating these minima from each other.
}
\label{RenPot}
\end{figure}
We show the potential in Killing coordinates in Fig. \ref{RenPot}. One can see that 
 these two inflationary plateaus are exact copies of each other: we have shown in Sec. \ref{Sec:Kil} that the potentials
 studied in \cite{Kallosh:2024ymt}  are symmetric with respect to the reflection symmetry $\tilde\vp \to -\tilde\vp$, and separately with respect to the reflection symmetry $\vt \to - \vt$. Similarly, the metric \rf{kinK}  is invariant with respect to the change $\tilde\vp \to -\tilde\vp$ and, separately,  $\vt \to - \vt$.  This means that the upper and the lower inflationary plateaus shown in Fig. \ref{RenPot} are {\it  physically equivalent} \cite{Kallosh:2024ymt}. 

Thus, inflation may begin not only at the plateau with $\tilde\vp > 0$, but also at the plateau with  $\tilde\vp < 0$ shown in Fig. \ref{RenPot}, and inflationary predictions do not depend on the choice of the plateau.

It is important to understand the relation between the potential shown in Fig. \ref{RenPot} and the potential in Figs.  \ref{Cartesian}-\ref{5band}. To guide the eye, one should pay attention to the red minima chain below the plateau's lower boundary in Figs. \ref{3band}, \ref{5band}. The centers of these minima are at $\tau_{1} = 0.5+ n$, where $n$ is any integer.  These minima are positioned along an infinitely long straight horizontal line  $\tau_{2} = \sqrt 3/2$. Meanwhile, in Killing coordinates, this infinite red chain's right and left sides bend upward, forming a red beaded necklace surrounding the upper plateau shown in Fig. \ref{RenPot}.

This fact becomes especially clear if one compares the contour plot in Fig. \ref{3Ridges} in the previous subsection with the contour plot of the potential  \rf{Renata2}  in Killing coordinates shown in the right panel of Fig. \ref{RenPot}. Thus, the upper plateau, including the red beaded necklace surrounding it, fully represents the infinite upper part of the half-plane with $\tau_{2} \geq \sqrt 3/2$ and $-\infty <\tau_{1} = \theta<  \infty$.

But what is the origin of its mirror image, the lower plateau, which is also surrounded by the red beaded chain? The answer is that it originates from the ridge at  $\vp <0$ shown in Fig. \ref{1Ridge}. This ridge is also surrounded by an infinitely long chain of red minima separated by saddle points. In Killing coordinates, the ridge's central part becomes as broad and flat as the upper plateau, and the red beaded chain surrounding the ridge becomes the necklace surrounding the lower plateau in Fig. \ref{RenPot}.

Let us confirm some of these observations by finding the positions of the centers of these red minima in new coordinates. The straight line of the minima with $\tau = 1/2+n +i  \sqrt 3/2$  in old coordinates shown in Figs. \ref{3band}, \ref{5band} is
$y=\sqrt 3/2$, $\theta=1/2 + n$.
In new coordinates $\tilde \vp, \vt$, this line of minima becomes a curve in with coordinates 
\be
\label{minima}
e^{\sqrt{2\over 3\alpha} \tilde\vp} \cos \Bigl(\sqrt{2\over 3\alpha} \vt \Bigr)= \sqrt 3/2 \ .  
\ee
This means that along this curve 
\be
\tilde\vp = \sqrt{3\alpha\over 2}  \ln \left({\sqrt{3}\over 2 \cos \big (\sqrt{2\over 3\alpha} \vt \big)}\right).
\ee
This explains why the chain of the red minima, which was a horizontal line in Cartesian coordinates, bends up close to the left and right sides of Fig. \ref{RenPot} where $\sqrt{2\over 3\alpha} \vt = \pm {\pi\over 2}$.

One can also track a line of the saddle points in Cartesian coordinates in Fig. \ref{5band} at $\tau=i+n$ and the corresponding curve of the saddle points in Killing coordinates at
\be\label{saddles}
e^{\sqrt{2\over 3\alpha} \tilde\vp} \cos \Big (\sqrt{2\over 3\alpha} \vt \Big)= 1 \ .
\ee
Finally, one may wonder how the hypersurface of the end of inflation looks in Killing coordinates. Unlike the inflationary predictions of $\alpha$-attractors, which are determined by the shape of the potential at the plateau, $V \approx V_{0}$, the exact position of the end of inflation at $V$ significantly smaller than $V_{0}$ is model-dependent. In the class of models that we study, the potential during inflation in these models is practically $\theta$-independent  \cite{Kallosh:2024ymt,KalLin2024}.  This helps to find the $\theta$-independent value of the field $\vp_{\rm end}$ where inflation ends, and the height of the potential $V_{\rm end}$ at that point. 

For the potential shown in Fig. \ref{1Ridge} with $\alpha = 2/3$, the slow-roll inflation ends at  $\tau_{2}  = e^{\vp} \approx 2$, $\vp \sim 0.7$. Therefore in Killing coordinates, the end of inflation is close to the line
\be\label{endofinfl}
e^{\sqrt{2\over 3\alpha} \tilde\vp}\cos \Big (\sqrt{2\over 3\alpha} \vt \Big) = 2 \  
\ee
for $\alpha = 2/3$.
As one can see, the red chain of the minima \rf{minima}, the chain of the saddle points  \rf{saddles},  and the end of inflation line \rf{endofinfl} closely follow each other. This means that most of the area above the red beaded neckless in Fig. \ref{RenPot} corresponds to the inflationary plateau.
\begin{figure}[H]
\centering
\includegraphics[scale=0.145]{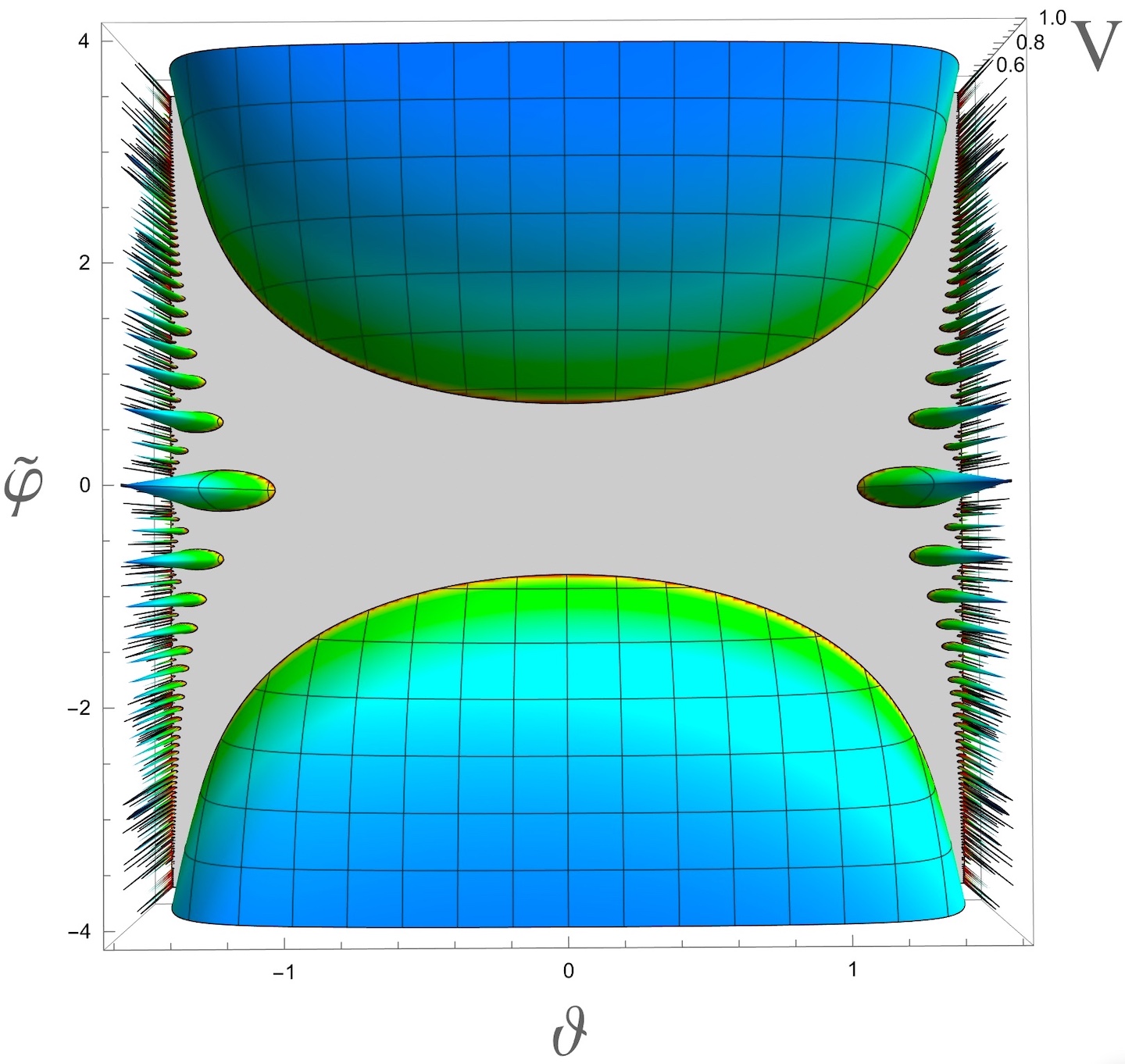}
\caption{\footnotesize The top view of the 3D plot of the upper part of the potential \rf{Renata2} for $V_{0} = 1$,  $n=1$, $3\alpha=2$,  in the coordinates \rf{Killing}  for $V \geq V_{\rm end} \sim 0.4 V_{0}$. It shows two inflationary plateaus where slow-roll inflation is possible.  The upper inflationary plateau is a mirror image of the lower one.}
\label{RenPot2}
\end{figure}
One can identify both inflationary plateaus explicitly by plotting the part of the potential higher than  $V_{\rm end} \sim 0.4V_{0}$, see Fig. \ref{RenPot2}.  Note that the upper inflationary plateau shown in Fig. \ref{RenPot2} represents not just the part of the fundamental domain with $-0.5<\theta<0.5$ and $V \gtrsim V_{\rm end}$, but the entire inflationary plateau with $-\infty <\theta<  \infty$, bounded from below by the end of inflation hypersurface with $V = V_{\rm end} \sim 0.4 V_{0}$, closely followed by the yellow-green chain of saddle points, and finally by the red beaded chain of minima of the potential with $V = 0$.

\section{Replication of inflationary plateaus} \label{replication}

In the previous section, using Killing coordinates \rf{Killing}, we found that the central ridge at $\vp < 0$ shown in Fig. \ref{1Ridge} is physically equivalent to the full upper part of the potential at $\vp > 0$, $-\infty < \theta < \infty$. It is as large as the plateau at $\vp >0$, and as flat and suitable for inflation. This is not entirely unexpected in the $SL(2,\mathbb{Z})$ invariant models in view of moduli invariance of the potentials and peculiarities of hyperbolic geometry, but it is still surprising, and surprises do not end there.

Indeed, in view of the periodicity of the potential with respect to the shift $\tau_{1} \to \tau_{1} +n$, {\it all} ridges beginning at $\tau_{1} = \theta= n$ (three and five of them are shown in Figs. \ref{3band}, \ref{5band} respectively) are physically equivalent to each other, and therefore {\it each of these ridges separately is physically equivalent to the full upper part of the potential supporting inflation \rf{Renata2}.}

This is just the beginning. Since each of the ridges is physically equivalent to the full upper part of the potential, one can find coordinates where {\it each of them} looks exactly like this upper part of the potential shown in Fig. \ref{5band}.  Let us demonstrate it. Some steps of this demonstration may seem somewhat tautological, but we will do it anyway.
\begin{figure}[H]
\centering
\includegraphics[scale=0.15]{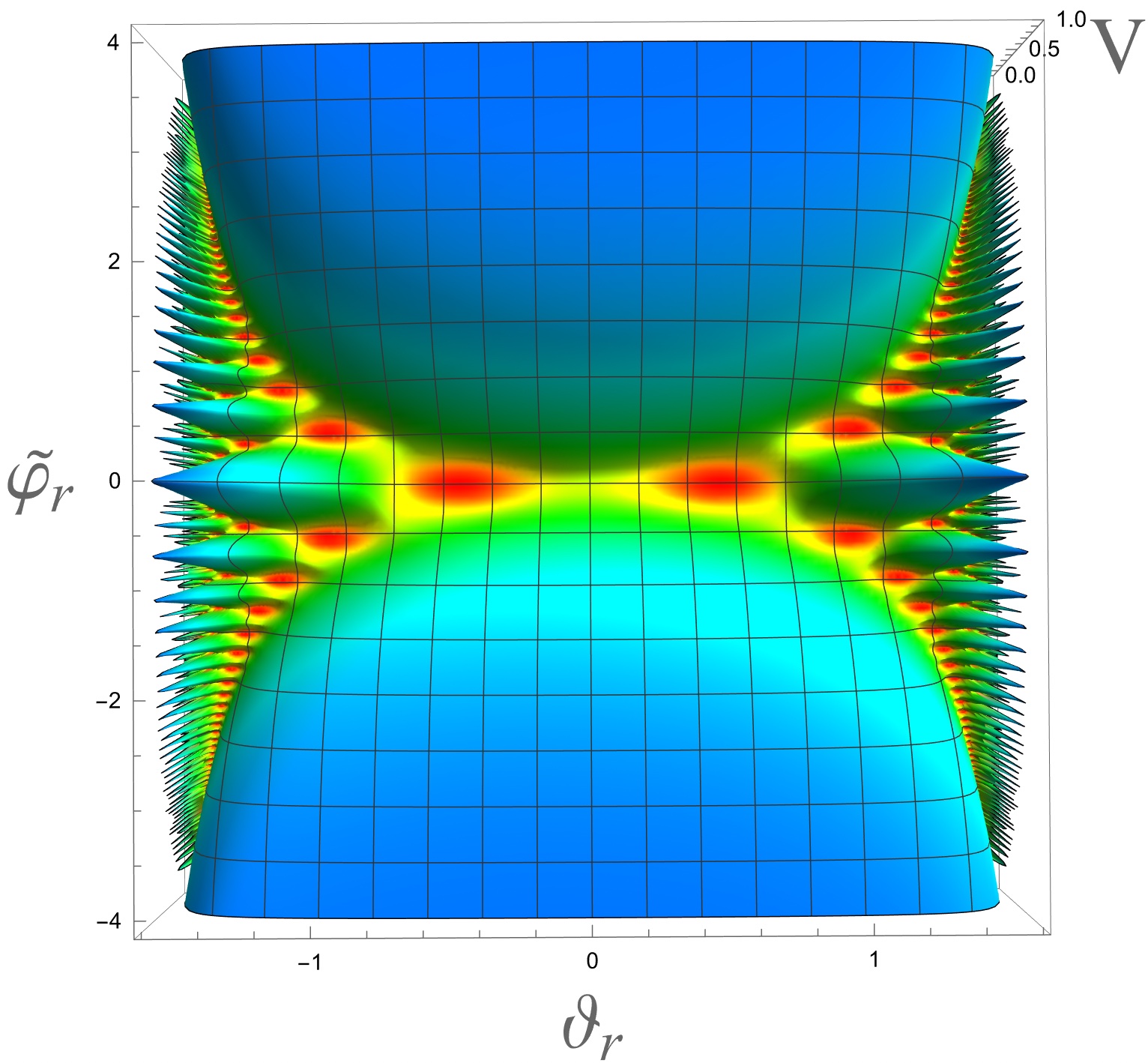}
\caption{\footnotesize The top view of the 3D plot of the upper part of the potential \rf{Renata2} for $V_{0} = 1$,  $n=1$, $3\alpha=2$,  in the new variables $\tilde\vp_{r}$ and  $\tilde\theta_{r}$. This potential coincides with the potential shown in Fig. \ref{RenPot}, but now its upper plateau corresponds to the  central ridge at $\vp < 0$ in 
Fig. \ref{5band}. }
\label{RenPotRidge}
\end{figure}
Killing coordinates $\tilde\vp$ and $\vartheta$, which we used to show the potential \rf{Renata2} in Fig. \ref{RenPot}, are related to the coordinates $\theta$ and $\vp$ by equations \rf{vp}, \rf{theta}. If we have any potential in Killing coordinates, one can always use these relations to express this potential in terms of $\theta$ and $\vp$. In particular, if desired, one can take the potential shown in Fig. \ref{RenPot} and return it back to its original form shown in Fig. \ref{5band}.

As we already mentioned, the potential shown in Fig. \ref{RenPot}, as well as the Killing metric, is invariant with respect to the reflection symmetry $\tilde\vp \to -\tilde\vp$, and with respect to the reflection symmetry $\vt \to - \vt$. Let us make a trivial change of variables $\tilde\vp \to -\tilde\vp_{r}$, $\tilde\vt \to -\tilde\vt_{r}$. 

Because of the invariance with respect to the change of variables $\tilde\vp \to -\tilde\vp_{r}$, $\tilde\vt \to -\tilde\vt_{r}$, the potential in the new coordinates, which we show in Fig. \ref{RenPotRidge}, looks exactly the same as the potential in Fig. \ref{RenPot}. However, its interpretation is different. Previously, the upper plateau with $\tilde\vp > 0$ surrounded by the red beaded chain corresponded to the upper part of the potential in Fig. \ref{5band}. But after the change of variables, the upper  plateau with $\tilde\vp_{r} > 0$ in Fig. \ref{RenPotRidge} corresponds to the central ridge at $\vp < 0$ in 
Fig. \ref{5band}.

As we already mentioned, one can use the relations \rf{vp} and \rf{theta} 
 \be\label{uu} 
 e^{\sqrt{2\over 3\alpha}  \vp }=   e^{\sqrt{2\over 3\alpha} \tilde\vp} \cos \Big (\sqrt{2\over 3\alpha} \vt \Big) \ , \qquad
 \theta=    e^{\sqrt{2\over 3\alpha} \tilde\vp} \sin \Big (\sqrt{2\over 3\alpha} \vt \Big) \ ,
 \ee
 to go from one set of coordinates to another and return back if desired. For example, one may start with the potential shown in Fig. \ref{RenPot} and return back to Fig. \ref{5band}. 

\begin{figure}[h!]
\centering
\includegraphics[scale=0.17]{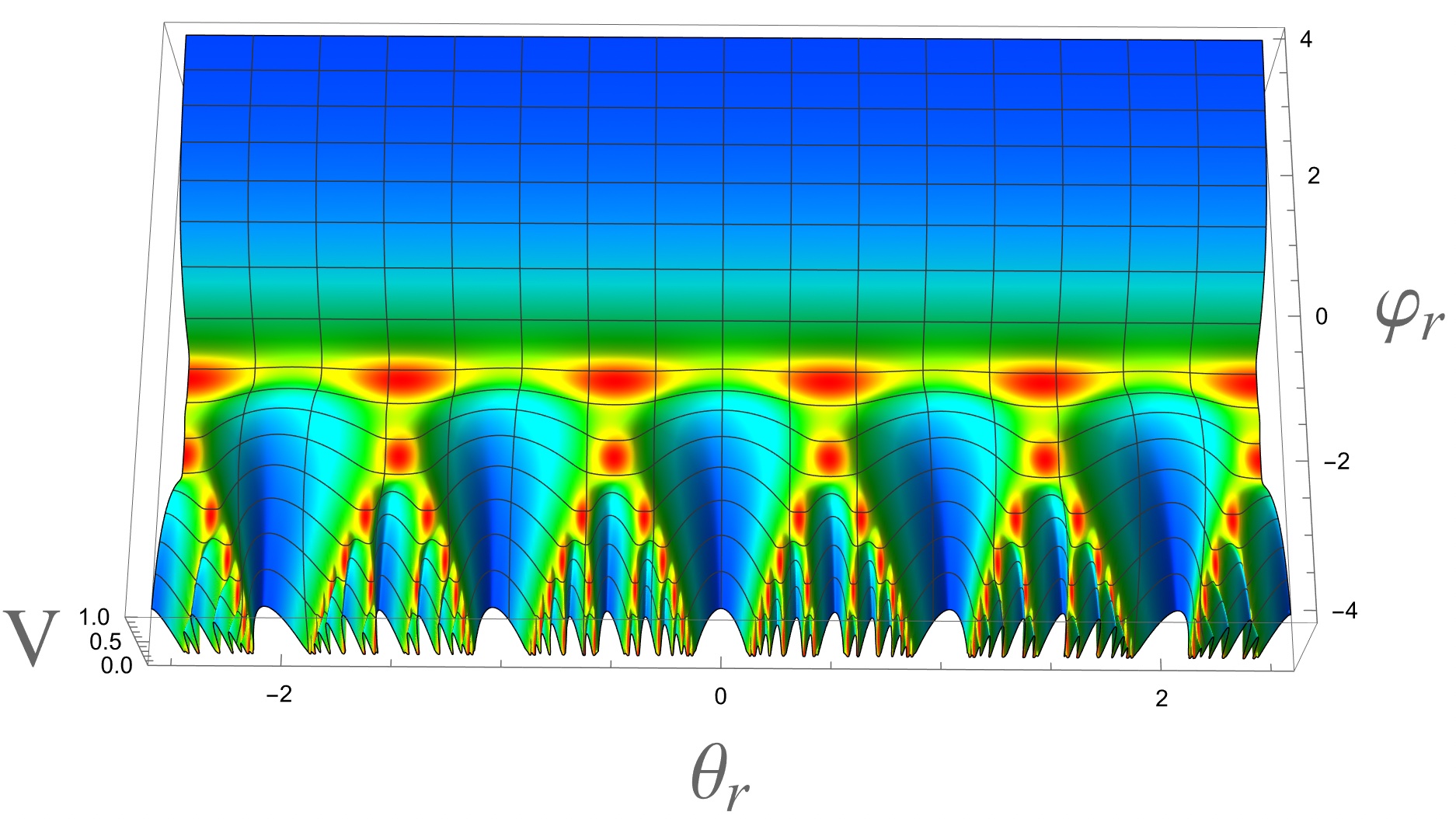}
\vskip -10pt
\caption{\footnotesize  Potential \rf{Renata2} as a function of the new variables $\tau_{r} = \theta_{r}+ i e^{\sqrt{2\over 3\alpha}  \vp_{r} }$. This potential looks exactly like the original potential shown in Fig. \ref{5band}, but its interpretation is very different, because the new variables  $\theta_{r}$ and $\vp_{r}$ appear as a result of a sequence of changes of variables discussed in the last two sections. In particular, the upper part of the potential in coordinates $\theta_{r}$ and $\vp_{r}$ describes the new Cartesian-type representation of what was previously shown as the upper plateau in Fig. \ref{RenPotRidge}, which was earlier shown as the lower plateau in Fig. \ref{RenPot}, which, in its turn,  what shown as a sharp ridge at the center of Figs. \ref{5band} and \ref{1Ridge}.  The large ridges in the lower part of this figure represent the smaller ridges surrounding the central ridge in Fig. \ref{1Ridge} in terms of the new variables  $\theta_{r}$ and $\vp_{r}$.}
\label{5band2}
\end{figure}

Now we are going to do the same for the new coordinates  $\vp_{r}$, $\theta_{r}$, using the same equations as Eq. \rf{uu}
  \be\label{vpr}
e^{\sqrt{2\over 3\alpha}  \vp_{r} }=   e^{\sqrt{2\over 3\alpha} \tilde\vp_{r}} \cos \Big (\sqrt{2\over 3\alpha} \vt_{r} \Big) \ , \qquad
\theta_{r}=    e^{\sqrt{2\over 3\alpha} \tilde\vp_{r}} \sin \Big (\sqrt{2\over 3\alpha} \vt_{r} \Big) \ .
\ee
Since the potential for the new variables $\tilde\vp_{r} $, $\vt_{r}$ in Killing coordinates coincides with the potential in variables $\tilde\vp$, $\vt$, and the transformation to the variables $\vp_{r}$ and $\theta_{r}$ is defined by the same rules as for $\vp$ and $\theta$, the resulting potential coincides with the one shown in Fig. \ref{5band},  but now it is the function of the new variables $\vp_{r}$, $\theta_{r}$ describing the ridge, see Fig. \ref{5band2}.

This confirms the expectations expressed at the beginning of this section: Since each of the ridges at $\vp < 0$ beginning at $\theta= n$ is physically equivalent to the full upper part of the potential, one can find a set of coordinates where any of these ridges looks exactly like the upper part of the potential shown in Fig. \ref{5band}.  We identified these coordinates and plotted the final result in Fig. \ref{5band2}.

One should emphasize that even though Fig. \ref{5band} and Fig. \ref{5band2} look identical and show two physically equivalent parts of the potential \rf{Renata2}, the interpretation of these two figures is very different. For example, the ridge at  $\theta = 0$, $\vp<0$ shown in Fig. \ref{5band} represents the upper part of the potential in Fig. \ref{5band2}. Similarly,  the ridge at  $\theta_{r} = 0$, $\vp_{r}<0$ shown in Fig. \ref{5band2} represents the upper part of the potential in Fig. \ref{5band}.

The difference between these two figures becomes especially important if, instead of the central ridge at $\theta_{r}= 0$, one considers the infinite family of ridges with $\vp_{r} < 0$ beginning at $\theta_{r}= n \not = 0$, as shown in Fig. \ref{5band2}.  
To understand the origin of these ridges, let us note that the horizontal red beaded chain in Fig. \ref{5band2} corresponds to the red beaded chain surrounding the lower plateau in Fig. \ref{RenPot}. There is a saddle point between each pair of these red minima, and small baby ridges touch these saddle points. 

In the new coordinates, in Fig. \ref{5band2}, these baby-ridges look as the ridges  $\theta_{r}= n \not = 0$ that we see now in Fig. \ref{5band2}, but they no longer look small in the new coordinates. In this coordinate system, in variables $\theta_{r}$ and $\vp_{r}$, all such ridges are copies of each other due to the periodicity in $\theta_{r}$. But now we know that the central ridge beginning at $\theta_{r}= 0$ is equivalent to the upper part of the potential shown in Fig.  \ref{5band}. This means that each of the small ridges initiating close to the central ridge in Fig. \ref{5band} is equivalent to the upper part of the potential shown in Fig.  \ref{5band}.

Thus, as a first step, we found that each large ridge beginning at $\theta = n$, $\vp<0$ is equivalent to the upper part of the potential shown in Fig. \ref{5band}. We will call these ridges progenitors.

 Then we developed a procedure demonstrating that each baby ridge originating close to one of the infinitely many progenitor ridges is also equivalent to the upper part of the potential shown in Fig. \ref{5band}.
 
This procedure is repeatable.  At each new step, we return to the same picture as the one shown in Fig. \ref{5band}, but in new coordinates. Each time, we see children, then grandchildren, then great-grandchildren of the progenitor ridges, and each time, one finds that these baby ridges are equivalent to the upper part of the potential shown in Fig. \ref{5band}. 

In short, the basic rule can be formulated as follows. Each ridge touching the upper part of the potential in variables $\theta$, $\vp$  (a progenitor) is equivalent to this upper part. Each ridge touching a progenitor is also equivalent to the upper part of the potential. By repeating this procedure over and over, one may conclude that {\it all}  ridges are physically equivalent to each other and, consequently, equivalent to the upper part of the potential containing an infinitely large inflationary plateau.

This general construction could be sufficient for our purposes, but we would like to have an alternative, more formal and systematic proof of the above statements.  In the subsequent sections, we will identify all saddle points of the potential. Then we will define a set of Killing coordinate systems with an origin at each saddle point. Manifest reflection symmetry $\Phi\to -\Phi$ will remain a property of the potentials in all Killing coordinate systems that we will introduce. As we will see, the new set of Killing coordinates will help us to represent any ridge of the potential as an inflationary plateau.

\section{Proliferation of saddle points}\label{Sec: fraction}

\subsection{Continued fractions}
Fractals
   are often associated with  continued fractions
\[
a_0 + \frac{1\kern6em}{\displaystyle
  a_1 + \frac{1\kern5em}{\displaystyle
    a_2 +\stackunder{}{\ddots\stackunder{}{\displaystyle
      {}+ \frac{1}{\displaystyle
        a_{n-1} + \frac{1}{a_n}}}
}}}
\]
Continued fractions of the regular type have all entries positive.
When we study the modular images in  $SL(2,\mathbb{Z})$ theories, we come across continued fractions of a more general nature. However, the feature of an endless proliferation of structures will be seen here via general continued fractions.

Consider the set of operations in Eq. \rf{nonun} which, according to  \cite{Rankin_1977}, represents  every element of $SL(2,\mathbb{Z})$, although it is a not unique. Suppose that we have started with some initial value of $\tau$, let us call it $\tau_0$. When all matrices in $T[q_0; q_1, \dots q_n]$ acted on $\tau_0$ a new coordinate $\tau_n [q_0; q_1, \dots q_n]
$ emerges, which is an image of $\tau_0$ depending on the operation $[q_0; q_1, \dots q_n]$ used to produce this image.
\be
[ U^{q_0} V U^{q_1} V\dots V U^{q_n}]  \tau_0 \to \tau_n [q_0; q_1, \dots q_n]\, , \qquad  q_i \in \mathbb{Z}\, , \qquad 0\leq i\leq n  \ .
\label{nonun1}\ee
Under the action of any element of $SL(2,\mathbb{Z})$ the relation between $\tau_n$ and  $\tau_0$ is of the form
\be
\tau_n = {a\tau_0 + b\over c\tau_0 +d} \ .
\label{taun}\ee
Here the integers $a,b,c,d$ depend on $[q_0; q_1, \dots q_n]$ and they are subject to condition $ad-cb=1$. This dependence can be studied using the explicit expressions for the matrices $U^{q_0} $ and $V U^{q_i}$, $i=1, \dots, n$.
\be
U^{q_0}= \left(\begin{array}{cc}1 & q_0 \\0 & 1\end{array}\right)\, , \qquad VU^{q_i} = \left(\begin{array}{cc}0 & -1 \\1 & \, q_i\end{array}\right) \ ,
\label{matrices}\ee
\be
U^{q_0} \tau = q_0 + \tau \, ,   \qquad VU^{q_{i}}  \tau= - {1\over q_i +\tau } \ .
\label{mat}\ee
The first shifts $\tau$ on an integer $q_0$, the second with $q_1$ adds an integer $q_1$ to the previous $\tau$ and makes an inversion, etc. If one starts with $\tau_0$ one finds after all steps in Eq. \rf{nonun1} that the image $\tau_n$ is
\be
\tau_n=
q_0 - \frac{1\kern6em}{\displaystyle
  q_1 - \frac{1\kern5em}{\displaystyle
    q_2 -\stackunder{}{\ddots\stackunder{}{\displaystyle
      {}- \frac{1}{\displaystyle
        q_{n-1} - \frac{1}{q_n+\tau_0}}}
}}}
\label{any}\ee
This is a generalized continued fraction, which is different from the regular continued fraction, where all sign and all $q_i$'s are positive. Here  we find that, as in the case of usual continued fraction, the results of the procedure in Eq. \rf{nonun1}, using Eq. \rf{mat} is always reduced to the form in 
Eq. \rf{taun}. One can also check   that for any specific choice of numbers in $[q_0; q_1, \dots q_n]$ the expression of the form  \rf{any} for any $n$, using Wolfram  is quickly evaluated and brought to the form \rf{taun}.

This procedure is particularly nice in case that we started with the first progenitor bridge between two minima at $\tau =i$. In such case
the continued fraction of the kind shown in \rf{any} for the proliferation of the saddle point has as an initial point $\tau_0=i$ in Eq. \rf{any}.
This expression can be simplified to
\be
\tau_n^{saddle} = {i a  + b\over i c +d}= {i  \over  c^2 +d^2}+  { bd+ac \over  c^2 +d^2} \ ,
\label{saddlen}\ee
where each of the integers $a,b,c,d$ depend on $[q_0; q_1, \dots q_n]$ and satisfy the constraint $ad-cb=1$.
In Cartesian coordinates both real and imaginary part of the new saddle points are always given by a rational number, being the ratio of some integers
\be
x_n [q_0; q_1, \dots q_n] =    { bd+ac \over  c^2 +d^2}\, , \qquad 
                            y_n [q_0; q_1, \dots
                        q_n]= {1  \over  c^2 +d^2} \ .
\ee

\subsection{Examples} 
\begin{enumerate}
  \item One step in \rf{nonun1} with $U^{q_0}$, $q_0=n$. This leads to 
  $a=1, d=1, c=0, b=n$ and 
  \be
 i \to  n + i 
 \ee
 This case is illustrated in Fig. \ref{5band} where we can see cases with $n=0, \pm 1, \pm 2$, so that there are total of 5 saddle points at $y=1, x=0, \pm 1, \pm 2$. Each of these saddle points has a plateau at values of $y\to \infty$ at $x=0, \pm 1, \pm 2$. At $y<1$ there is also something like a plateau, so the saddle point makes a bridge, however, these all become sharp ridges very soon.
 
 In Figs. \ref{RenPot} the line $y=1$ takes a curved form $e^{\sqrt{2\over 3\alpha} \tilde\vp} \cos \Big (\sqrt{2\over 3\alpha} \vt \Big)=1$. It surrounds the plateaus for both positive and negative $\tilde \vp$. In the conterplot there are related curves. Many saddle points  can be seen in all these figures but not the ones which are at some distance from the plateaus.

  \item Next example is with  generator $U^{q_0}V U^{q_1}  $, $q_0=0$ and  $q_1=\pm n$. Using Eq. \rf{matrices} we find new saddle points are at 
  $\tau^{saddle} = \pm {n\over n^2+1}+ {i\over n^2+1} $. Specific cases in this group are, for $n=1,2,3$
  \be
  \tau^{saddle} = \pm {1\over 2}+ {i\over 2}\, , \quad   \tau^{saddle} = \pm {2\over 5}+ {i\over 5}\, , \quad   \tau^{saddle} = \pm {3\over 10}+ {i\over 10} \ .
\label{case1}  \ee
\begin{figure}[H]
\centering
\includegraphics[scale=0.15]{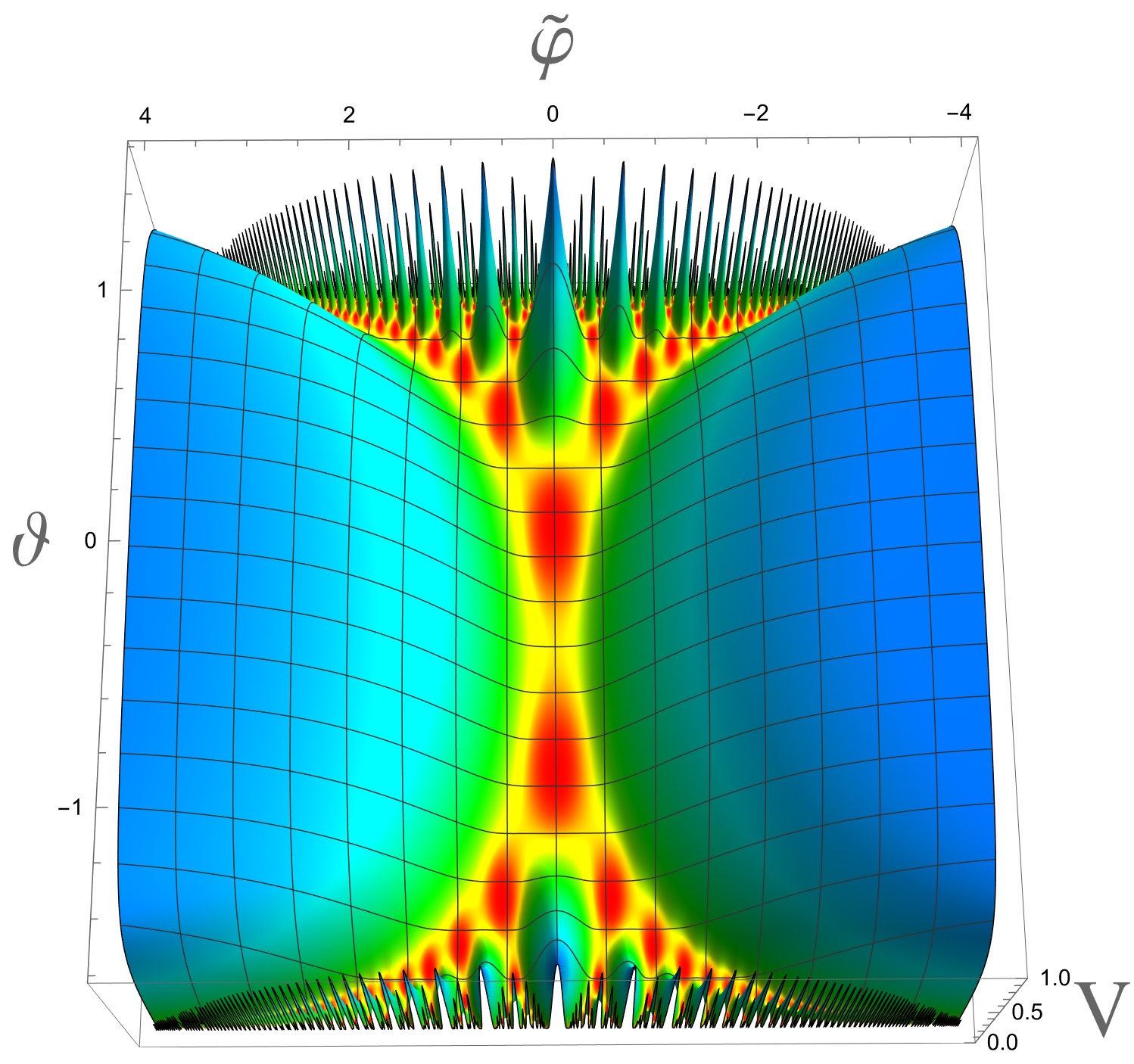}
\vskip -10pt
\caption{\footnotesize   Same potential \rf{Renata2} as in Fig. \ref{RenPot} in Killing coordinates but with the better view on proliferation of the saddle points. At $\Phi=\tilde \vp=\vt=0$ there is a saddle point at $\tau =i$. The saddle points keep doubling when moving away from $\Phi=0$.}
\label{saddles1}
\end{figure}
  
 \item Next example is with  generator $U^{q_0}V U^{q_1} V U^{q_2}$, $q_0=0$ and  $q_1=\pm k$, $q_2= \mp 1$. Using Eq. \rf{matrices} we find new saddle points are at 
  $\tau^{saddle} = \pm {2k+1\over 2k^2+2k +1}+ {i\over 2k^2+2k +1} $.  Specific cases in this group are, for $k=1,2,3$
  \be
  \tau^{saddle} = \pm {3\over 5}+ {i\over 5}\, , \quad   \tau^{saddle} = \pm {5\over 13}+ {i\over 13}\, , \quad   \tau^{saddle} = \pm {7\over 25}+ {i\over 25} \ .
\label{case2}  \ee
\end{enumerate}

Related examples of the proliferation of the saddle point images were given in \cite{Schimmrigk:2021tlv}. Here our examples  are shown to be special cases of the general formula in Eq. \rf{nonun1} which describes all possible saddle points.
This proliferation of saddle points in Killing coordinates in \rf{Kil} is also shown in Fig. \ref{saddles1}.

An interesting feature we can see, for example, in Figs. \ref{Cartesian}, \ref{5band} is that the saddle point at $\tau=i+n$ is a bridge between a plateau and a ridge. We can see now, in the Killing coordinate system in Fig. \ref{RenPot},  that this same saddle point at $\tau=i$ is a bridge between two plateaus. The next level saddle point behind each minimum in Fig. \ref{5band} which is at $\tau =\pm{1\over 2} + i{1\over 2}$ is a bridge between two minima. There is a plateau behind one of them and a ridge behind the other. 
We will see later, in the Killing coordinates systems in Fig. \ref{17}, that this same saddle point at $\tau =\pm{1\over 2} + i{1\over 2}$ is a bridge between two minima, each one having a plateau descending to these minima.

\section{Proliferation of  plateaus  in new Killing coordinate systems }\label{Sec: prolif}

\subsection{Continued fractions, saddle points, and a new family of Killing coordinate systems}

Looking at Fig. \ref{5band} in Cartesian coordinates, it is hard to imagine that all the ridges at $y= e^{\sqrt{2\over 3\alpha}  \vp } <1$, i.e. $\vp<0$ are not really ridges but plateaus. But we know that the actual distance between the points is $ds^2= {dx^2+dy^2\over y^2}$.  Therefore, the physical distance grows when we approach the half-plane boundary at $y=0$, but the Cartesian and axion-inflaton coordinates do not reflect it. That is why in these coordinates the ridges look sharper and sharper close to the boundary.

We have also seen  that by changing the Cartesian coordinates to Killing coordinates, we can see the 
very first ridge in Fig. \ref{5band} which is across the plateau at $y>1$ at about $\theta=0$ in the Killing coordinates system 
in Fig. \ref{RenPot} becomes an exact image of the plateau in the Killing coordinates.

But can we find a way to see that all other sharp ridges in Fig. \ref{5band}, which are close to the line $y=1$ and the ones behind it, which form the ever-doubling trees closer and closer to the  $y=0$  boundary, are actually all plateaus? The answer to this question is positive. The procedure we proposed and described in this paper is valid for an infinite number of ridges. It is based on the properties of Killing coordinates where the distances between various points are $\tilde \vp$ independent. Moreover,   at small $\vt=0$  away from the boundary $\sqrt{2\over 3\alpha} \vt \to \pm {\pi\over 2}$ both coordinates are canonical, and the potential shows all distances not distorted by a non-trivial geometry.

Let us start  with the original form of Killing coordinates  in \rf{original}
\be
\tau = i e^{\Phi} =  i e^{\sqrt{2\over 3\alpha} (\tilde \vp -i \vt)}\, , \qquad -\infty <\tilde \vp < \infty \, , \qquad   -\pi/2 < \sqrt{2\over 3\alpha} \vt < \pi/2
\label{K}\ee
with the inflaton $\tilde \vp$-independent metric in Eq. \rf{kinK}. 

{\it Our goal now is to find new Killing coordinate systems where any specific saddle point is at the center of the corresponding coordinate system}. We have derived the general formula describing the proliferation of saddle points using continued fractions in Appendix \ref{Sec: fraction} and found that after a set of $SL(2, \mathbb{Z})$ transformations defined in Eq. \rf{nonun} the new saddle point is at 
\be
\tau_n^{saddle} =  {i  \over  c^2 +d^2}+  { bd+ac \over  c^2 +d^2} \ ,
\label{nsaddle}\ee
where each of the integers $a,b,c,d$ depend on $n+1$ numbers $[q_0; q_1, \dots q_n]$ and satisfy the constraint $ad-cb=1$.
Note that the positions of all saddle points are always given by rational numbers, being ratios of integers in Eq. \rf{nsaddle}, where $y_n={1  \over  c^2 +d^2}$ and $x_n={bd+ac \over  c^2 +d^2}$.

We can start by  perform a  general  $SL(2,\mathbb{R})$ transformation on  $\tau$ 
\be
\tau' =  {\a \tau +\b\over \g \tau + \d} , \qquad \a,\b,\g,\d \in \mathbb{R} \ ,
\label{R}\ee 
which preserves the kinetic term. For $\tau =ie^{\Phi}$ this will generate a new coordinate system
\be
\tau'(\Phi; \a,\b,\gamma,\delta)=   {i\, \a  e^{\Phi} +\b\over i\, \gamma   e^{\Phi} + \delta} \ .
\label{phi1}\ee
To simplify it we make a choice
 $\gamma=0\, ,  \a \delta=1$ and get
\be
\tau'|_{\gamma=0\, , \, \a  \delta=1}=  i {\,  e^{\Phi} \over  \delta^2} +{\b\delta \over \delta^2} \ .
\label{phi11}\ee
Now we observe that if we take as a new Killing coordinate system an expression 
\be
\tau^+ [a,b,c,d]= i {e^{\Phi}\over c^2+d^2} + { bd+ac \over  c^2 +d^2}
\label{CSI}\ee
we have reached the goal: at $\Phi=0$, at the center of a new coordinate system,  it is a saddle point of our choice \rf{nsaddle} under the condition that our general $SL(2,\mathbb{R})$ transformation \rf{R} is actually a 
 {\it restricted set of  $SL(2,\mathbb{R})$ transformation} on the modulus, where
\be
\gamma=0\, , \qquad \a  \delta=1\, , \qquad
\delta^2= c^2+d^2\, , \qquad \beta \delta = bd+ac \ .
\label{constr}\ee
Thus, $\a, \b, \gamma, \delta$ are not general $SL(2,\mathbb{R})$ transformation which would require $\a, \b, \gamma, \delta \in \mathbb{R}$. They are {\it restricted  $SL(2,\mathbb{R})$ transformation} satisfying a set of constraints in Eq. \rf{constr}.

We could have started with the coordinate system 
\be
\tau = i e^{-\Phi} =  i e^{\sqrt{2\over 3\alpha} (-\tilde \vp +i \vt)}\, , \qquad -\infty <\tilde \vp < \infty \, , \qquad   -\pi/2 < \sqrt{2\over 3\alpha} \vt < \pi/2 \ ,
\label{Km}\ee
since the inversion symmetry $\Phi\to -\Phi$ is a property of our potentials in the original Killing coordinate system. In such case, after analogous {\it restricted  $SL(2,\mathbb{R})$ transformation} we would find new Killing coordinate systems of the form
\be
\tau^- [a,b,c,d]= i {e^{-\Phi}\over c^2+d^2} + { bd+ac \over  c^2 +d^2} \ .
\label{CSIm}\ee
The original Killing coordinate system has the symmetry under $\Phi \to - \Phi$ and a stronger case
\be
\tilde \vp=- \tilde \vp
\ee
and independently
\be
\vt \to - \vt 
\ee
The new Killing coordinate systems  produced by {\it restricted  $SL(2,\mathbb{R})$ transformation} are
\be
\tau^{\pm} [a,b,c,d]= i {e^{^{\pm} \Phi}\over c^2+d^2} + { bd+ac \over  c^2 +d^2} \ .
\label{CS1}\ee
Now we would like to find out how these new coordinate systems $\tau^+ [a,b,c,d]$ and $\tau^- [a,b,c,d]$ are related to each other.

To find out, we can 
 choose $\delta=0\, ,  \gamma \beta=-1$ in Eq. \rf{phi1} and get
\be
\tau'|_{\delta=0\, , \, \gamma \beta=-1}=   {i\,   e^{-\Phi} \over  \gamma^2} +{\a\gamma \over \gamma^2} \ .
\label{phi11a}
 \ee
Now we have to request that
\be
\delta=0\, , \qquad \gamma  \beta=-1\, , \qquad
\gamma^2= c^2+d^2\, , \qquad \a \gamma = bd+ac
\label{constrm}\ee
Thus, to relate $\Phi$ to $-\Phi$ we have to request that also
\be
 \delta^2\to \gamma^2 \, , \qquad  \b\delta \to \a\gamma \ .
\ee
We can describe the first choice and the second choice of $\a, \b, \g,\d$ with the following  $SL(2,\mathbb{R})$ matrices
\be
I=\left(\begin{array}{cc}\alpha & \beta \\0 & \delta\end{array}\right) \, , \qquad \qquad II=\left(\begin{array}{cc}-\beta & \alpha \\0 & \gamma\end{array}\right)\ . \ee
At $\gamma= \delta$  between these two matrices there is an $SL(2,\mathbb{Z})$ transformation 
\be
I= \Omega \, II \, , \qquad \Omega=\left(\begin{array}{cc}0 & 1 \\-1 & 0\end{array}\right) \ .
\label{Omega}\ee
And since we compare \rf{CSI} with \rf{CSIm} it means that in our case 
\be
\delta^2= \gamma^2=c^2+d^2\, , \qquad \beta \delta = \alpha \gamma =bd+ac \ .
\ee
Thus we have found that the relation between our two Killing coordinate systems in \rf{CSI} and in  \rf{CSIm} is an $SL(2,\mathbb{Z})$ transformation \rf{Omega}
\be
\Phi \to -\Phi  \qquad \Rightarrow \qquad SL(2,\mathbb{Z}) \ .
\ee
The potential in the coordinate system in \rf{CSI} is equal to the one in \rf{CSIm}, i.e., in Killing coordinate systems and the symmetry $\Phi \to -\Phi $ takes place in all cases in \rf{CSI} and \rf{CSIm}.  It is interesting that both coordinate systems at $\Phi=0$ always reproduce the saddle point proliferation formula \rf{nsaddle}.

Therefore, using the new Killing coordinate systems (which are restricted $SL(2,\mathbb{R})$ but not $SL(2,\mathbb{Z})$ images of the original one), we will produce our $SL(2,\mathbb{Z})$  invariant potentials which are different from the ones in the original Killing coordinate system 
\rf{K}. In particular, these new coordinate systems are designed to have at $\Phi=0,  \tilde \vp=\vt=0$ a saddle point defined in Eq. \rf{nsaddle}  where each of the integers $a,b,c,d$ depend on $[q_0; q_1, \dots q_n]$ and satisfy the constraint $ad-cb=1$. 

The case of the original Killing coordinate system $\tau = i e^{\Phi}$ is the case where  in $\tau_n^{saddle} = {i a  + b\over i c +d}$ we have
$a=d=1, b=c=0$. In this way, making new and new choices of $a,b,c,d$ via new and new choices of $[q_0; q_1, \dots q_n]$, we will see the new form of our $SL(2,\mathbb{Z})$ invariant potentials which at $\Phi=\tilde \vp=\vt=0$ have a saddle point of our choice, away from $y=1$ line, closer and closer to the boundary $y=0$. The corresponding 3D plots of the potentials in new Killing coordinate systems will present the view of surrounding mountains from the generic saddle point closer and closer to the boundary $y=0$.

 Let us summarize the properties of our  new Killing coordinate systems \rf{CSI} and its $SL(2,\mathbb{Z})$ image in 
\rf{CSIm} 
\begin{itemize}
  \item All of them have the same kinetic term given in Eq. \rf{kinK}
  \item Since they represent $SL(2,\mathbb{R})$ images of the original Killing coordinate system, described in this section, {\it the potential in these new coordinate systems is different from the potential in the original Killing coordinate system}.
  \item At $\Phi=0$, $\tilde \vp=\vt=0$ each new coordinates system will have one of an infinite numbers of the $SL(2,\mathbb{Z})$ images of saddle points at the center.
  \item The potential in all coordinate systems in \rf{CSI} is invariant under $\Phi\to -\Phi$ due to $SL(2,\mathbb{Z})$ symmetry.
\end{itemize}
 \subsection{Examples}

Our examples of Killing coordinate systems in Eq. \rf{CS1} are
\be
\tau^n = {i e^{\Phi} \over n^2+1}  \pm {n\over n^2+1}  \ .
\ee

At $n=0$, it is the original one in Eq. \rf{K}. At $n=1,2,3$
 \be
  \tau^{n=1} = \pm {1\over 2}+ {i e^{\Phi}\over 2}\, , \quad   \tau^{n=2}  = \pm {2\over 5}+ {i e^{\Phi}\over 5}\, , \quad   \tau^{n=3}  = \pm {3\over 10}+ {i e^{\Phi}\over 10} \ .
\label{case1CS}  \ee
At $\Phi=0$, the potential in these coordinate systems has saddle points given in Eq. \rf{case1}.

\begin{figure}[H]
\centering
\includegraphics[scale=0.12]{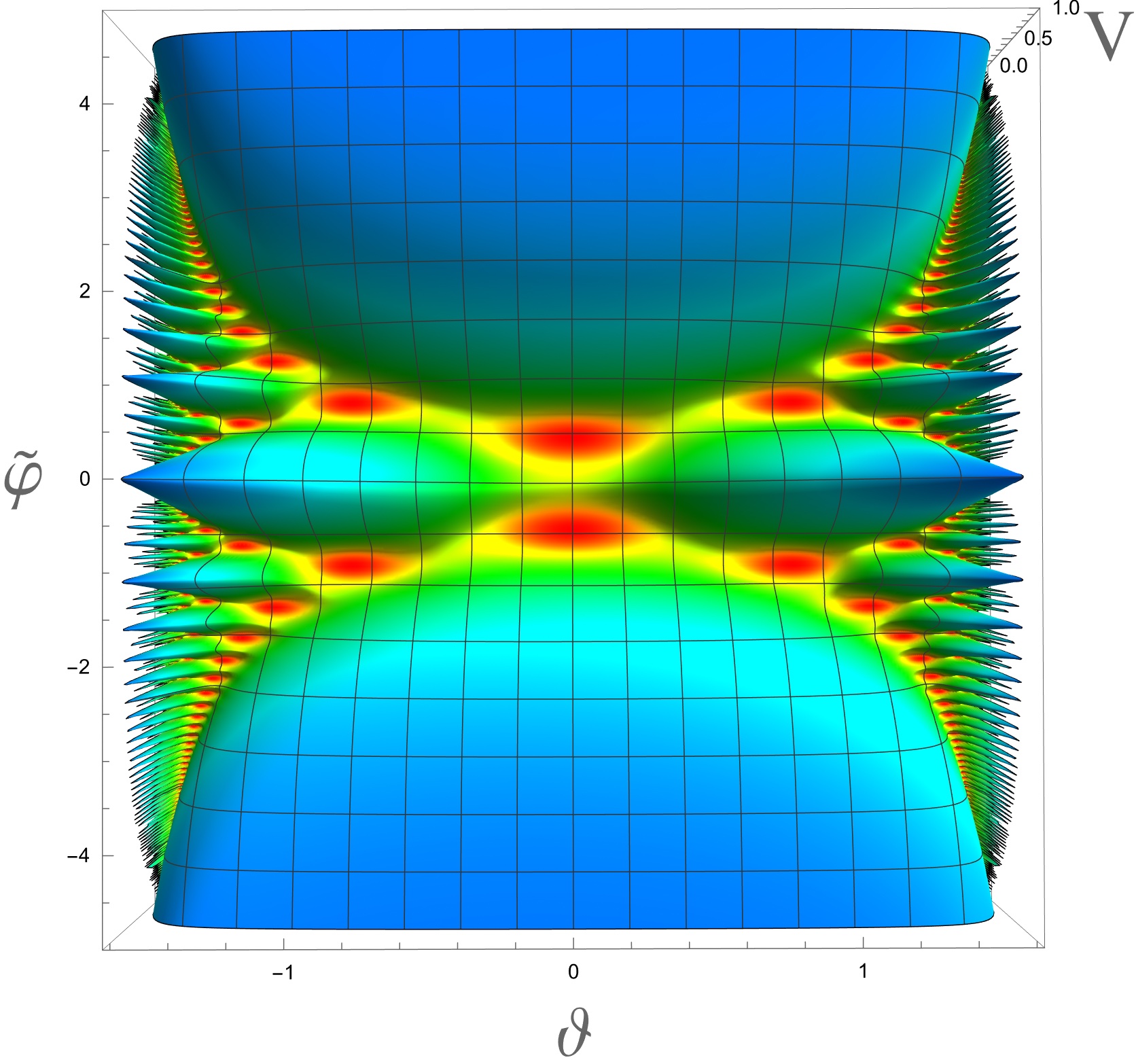} \hskip 0.5 cm \includegraphics[scale=0.12]{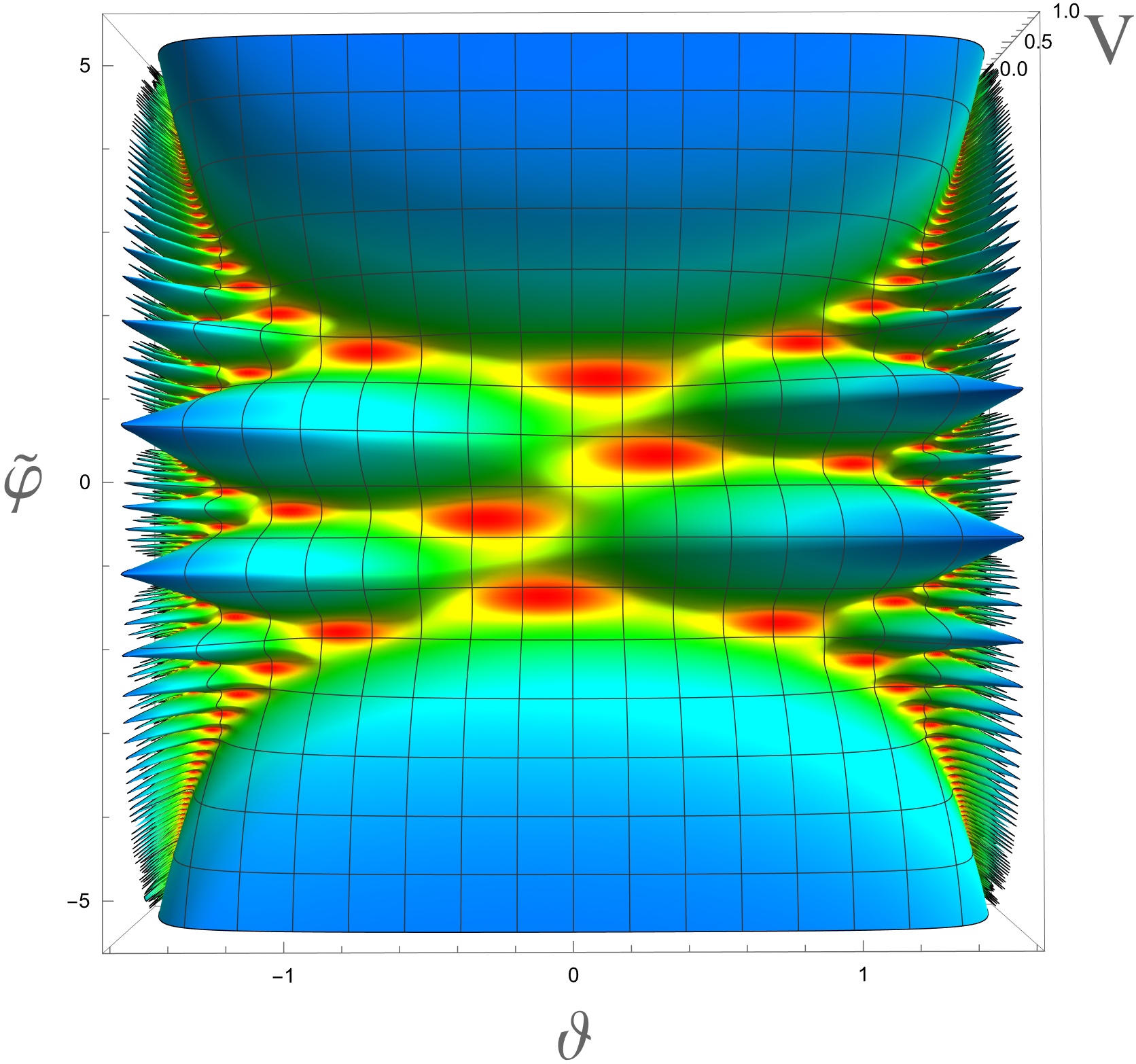}
\caption{\footnotesize  [Left panel:] The same potential \rf{Renata2} as in Figs. \ref{5band}, \ref{RenPot},  but    in Killing coordinates \rf{CS1}  with $n=1$.
At $\Phi=\tilde \vp=\vt=0$ there is a saddle point at $\tau = {1\over 2}+ {i \over 2}$. This saddle point between one minimum on each side is surrounded by two plateaus now. The lower plateau in this figure was looking like a sharp ridge in Fig. \ref{5band}. [Right panel:]   Potential \rf{Renata2}    in Killing coordinates $\tau^n = {i e^{\Phi} \over n^2+1}  + {n\over n^2+1}$  with $n=2$.
At $\Phi=\tilde \vp=\vt=0$ there is a saddle point at $\tau =\pm {2\over 5}+ {i \over 5}$. This saddle point between two minima on each side is surrounded by two plateaus now. The lower plateau was a sharp ridge in Fig. \ref{5band}, different from the ridge corresponding to the plateau shown in the left panel of this figure.}
\label{17}
\end{figure}

Fig. \ref{17} shows the potential in these coordinates for $n = 1$ (left panel) and $n = 2$ (right panel). One can see that between the two plateaus in the left panel of Fig. \ref{17}, there is one saddle point and two minima near $\tilde \vp=\vt=0$, as we saw before in Fig. \ref{5band}. But now, in the left panel of Fig. \ref{17}, there are two plateaus descending to these minima, from above and from below.  The right panel of  Fig. \ref{17} corresponds to the case $n = 2$.  One can see that between the two plateaus in this figure, there is one saddle point at $\tilde \vp=\vt=0$,  and two minima on the way to each plateau,  as we saw before in Fig. \ref{5band}. In the right panel of Fig. \ref{17}, there is a symmetry $\Phi\to -\Phi$, but there are no separate reflection symmetries for $\tilde \vp$ and $\vt$.

\begin{figure}[H]
\centering
\includegraphics[scale=0.12]{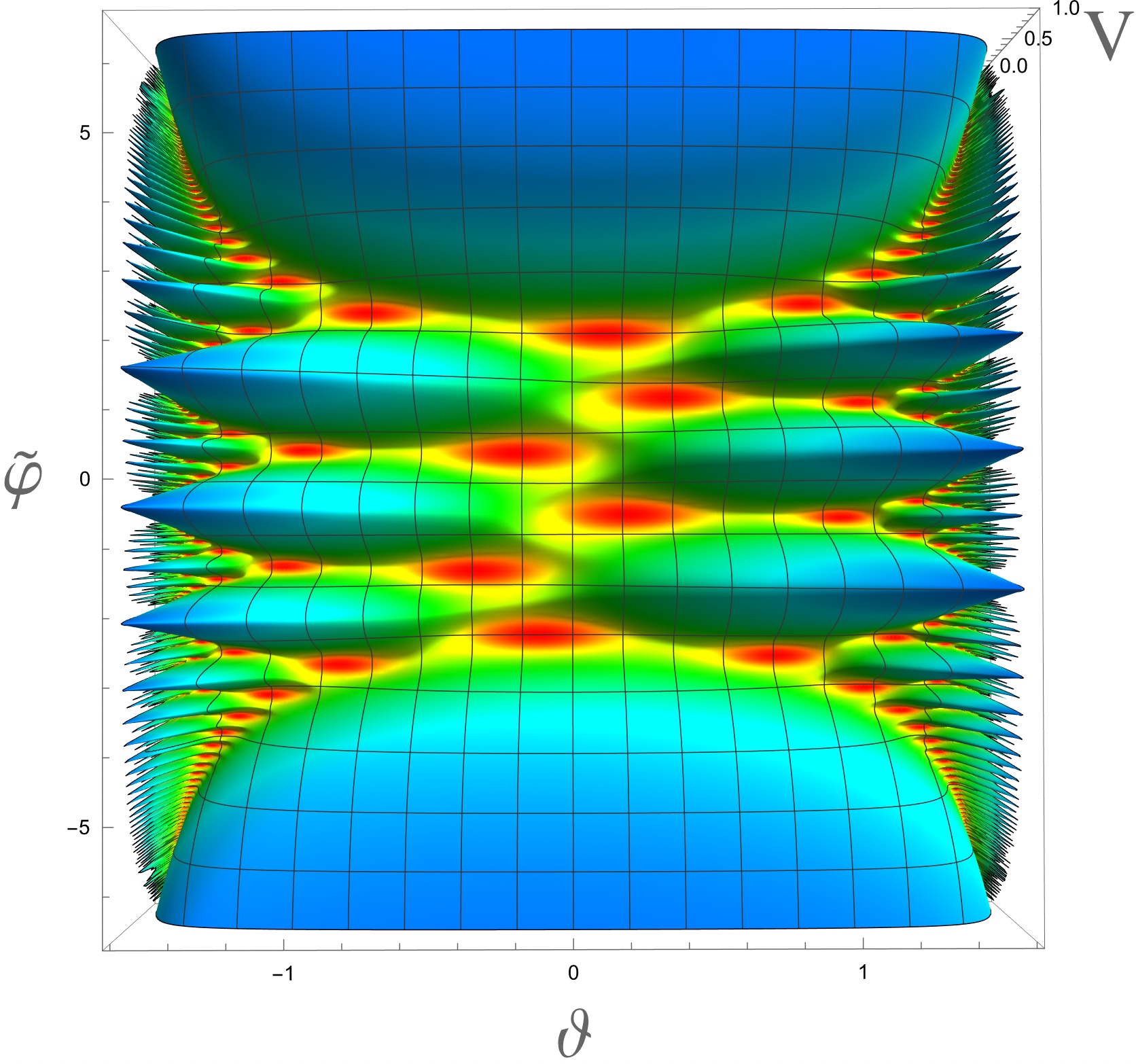} \hskip 1cm \includegraphics[scale=0.12]{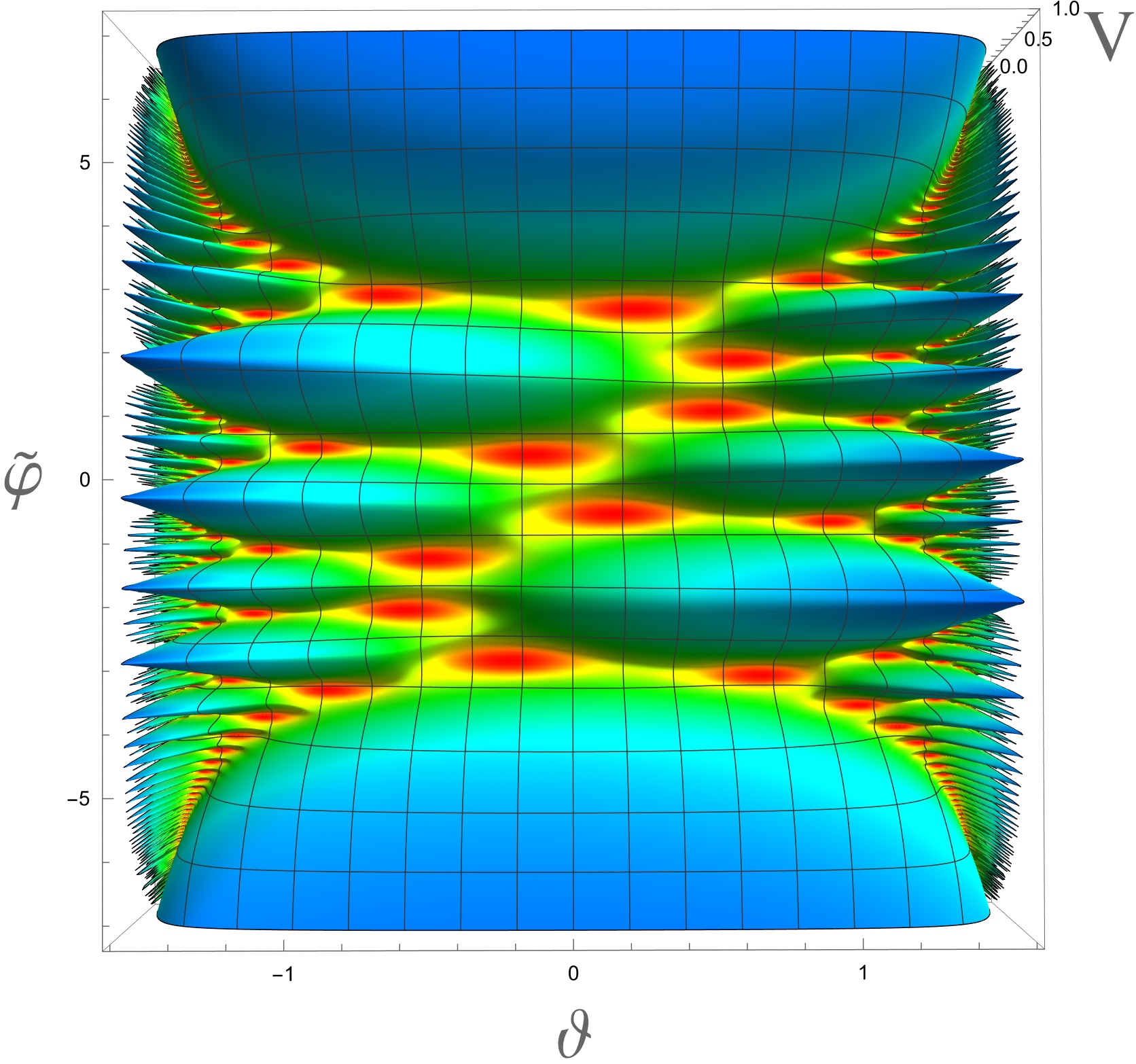}
\vskip -6pt
\caption{\footnotesize  [Left panel:]  Potential \rf{Renata2}    in Killing coordinates $\tau^k = {i e^{\Phi}\over 2k^2+2k +1}   + {2k+1\over 2k^2+2k +1}$  with $k=2$.
At $\Phi=\tilde \vp=\vt=0$ there is a saddle point at $\tau =\pm {5\over 13}+ {i \over 13}$.    [Right panel:] Potential \rf{Renata2}    in Killing coordinates $\tau_{Killing}^k = {i e^{\Phi}\over 2k^2+2k +1}   + {2k+1\over 2k^2+2k +1}$  with $k=3$. At $\Phi=\tilde \vp=\vt=0$ there is a saddle point at $\tau = {7\over 25}+ {i \over 25}$.  }
\label{k2}
\end{figure}

\begin{figure}[H]
\centering
\includegraphics[scale=0.51]{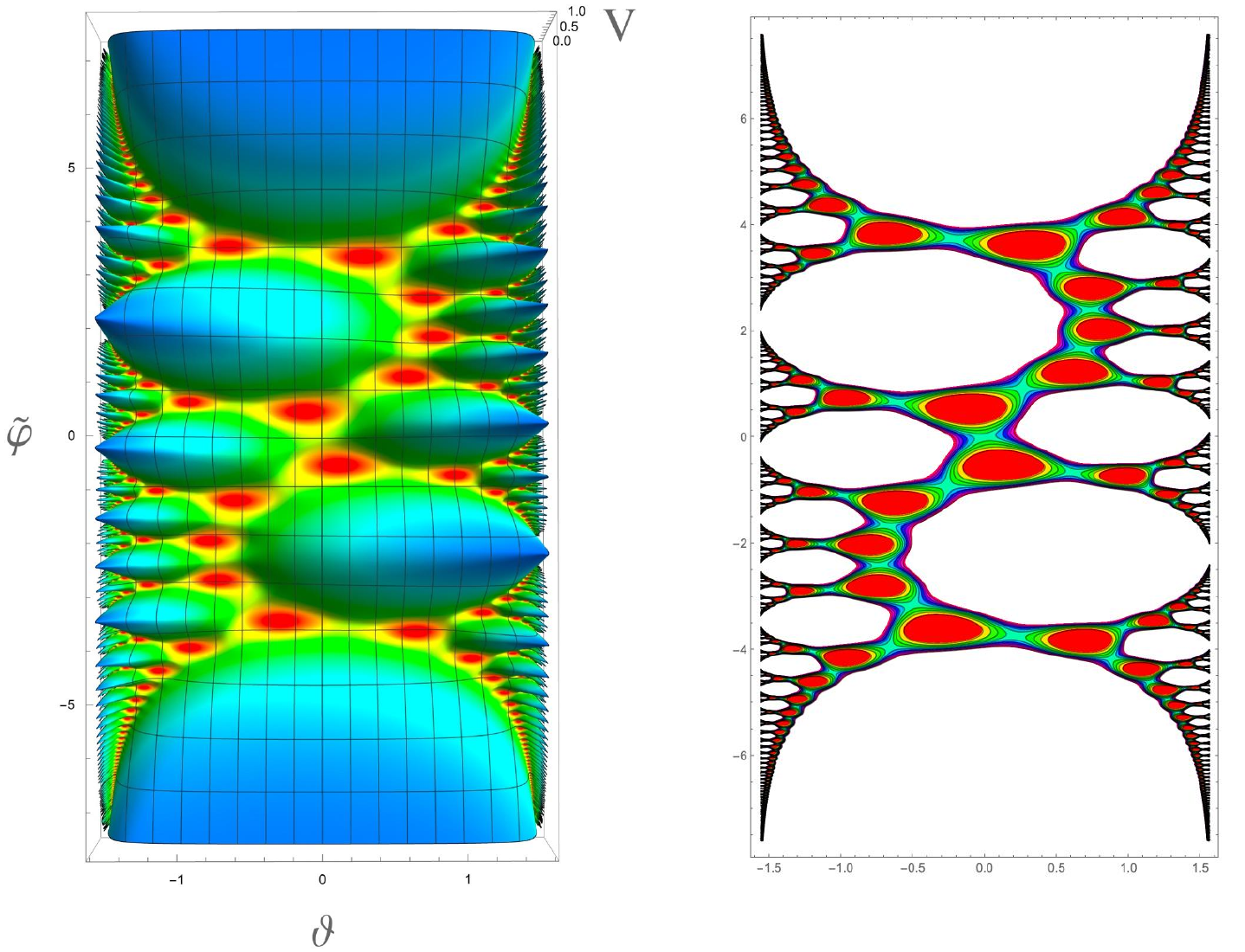}
\vskip -10pt
\caption{\footnotesize [Left panel:]  Potential \rf{Renata2}    in Killing coordinates $\tau_{Killing}^k = {i e^{\Phi}\over 2k^2+2k +1}   + {2k+1\over 2k^2+2k +1}$  with $k=4$. At $\Phi=\tilde \vp=\vt=0$ there is a saddle point at $\tau = {9\over 41}+ {i \over 41}$.  This saddle point between five minima on each side is surrounded by two plateaus now. The lower plateau looked like a sharp ridge in Fig. \ref{5band}, but in the new coordinates, it becomes an exact copy of the upper plateau, up to the $SL(2,\mathbb{Z})$ transformation $\Phi\to -\Phi$.  [Right panel:] The contour plot of the lower part of the potential at $V < 0.1\, V_{0}$ exhibits a fractal distribution of minima and saddle points. Red areas surround the minima of the potential; green areas correspond to the saddle points. }
\label{long}
\end{figure}

 Another branch of the proliferation of saddle points defines a new set of coordinate systems 
 \be
\tau^k ={i e^{\Phi}\over 2k^2+2k +1}   \pm {2k+1\over 2k^2+2k +1} \ .
\ee
At $k=0$, it is the original one in Eq. \rf{K}. At $k=1,2,3,4$ at $\Phi=\tilde \vp=\vt=0$ there are a saddle points at ${i e^{\Phi}\over 2k^2+2k +1}   \pm {2k+1\over 2k^2+2k +1}$ 
\bea
&&  \tau^{k=1} = \pm {3\over 5}+ {i\over 5}, \quad   \tau_{Killing}^{k=2} = \pm {5\over 13}+ {i\over 13}, \cr
&&   \tau^{k=3} = \pm {7\over 25}+ {i\over 25}, \quad  \tau_{Killing}^{k=4} = \pm {9\over 41}+ {i\over 41} \ .
\label{case2CS}  \eea
We show the corresponding potentials in Figs. \ref{k2} (left and right panels) and in Fig. \ref{long}. 

One can see that between the two plateaus in the left panel of Fig. \ref{k2}, there is a saddle point at the center and 3 minima on the way to each plateau, and between the two plateaus in the right panel of  Fig. \ref{k2}, there is a saddle point at the center and 4 minima on the way to each plateau. 

Finally, in Fig. \ref{long} we show the potential for $k = 4$. There are 10 minima on the way from the upper plateau to the lower plateau, i.e.,  5 minima on each side on the saddle point at the center. With each step with an increasing number of the minima separating the plateaus, the distance between the upper and lower plateaus increases. We did not show it in Figs. \ref{17} and \ref{k2}, compressing the images to the same square frame, but we show this distance increase in Fig. \ref{long}. In addition,  we show there the fractal distribution of minima and saddle points of the potential in these coordinates for $k = 4$.

Note that there is a saddle point at the center of each of these figures, but these saddle points have different positions in Fig. \ref{5band}. Similarly, there is a large plateau at the bottom of each figure, which is an exact copy of the plateau at the top of each figure. The upper plateau in each of these figures shows the entire upper part of the half-plain shown in Fig. \ref{5band}, whereas the lower plateaus in each of these figures correspond to different ridges, each of them being an exact copy of the upper ridge.

Thus, we have found that there are choices of coordinate systems that have an ever-increasing position of the saddle point at the center $\tilde \vp= \vt=0$. From this position, one can see two plateaus supporting inflation at all sufficiently large values of $|\tilde\vp|$. As in all previous examples, the upper plateau is exactly equivalent to the lower one because the potential and the metric are invariant under the transformation $\Phi\to -\Phi$ due to $SL(2,\mathbb{Z})$ symmetry.  Fig. \ref{long} shows a beautiful pattern of symmetry under $\tilde \vp$ reflection simultaneously with $\vt$ reflection.

The examples given above illustrate our general method of finding a coordinate system where the equivalence between ridges and plateaus becomes manifest.

\section{Summary and Discussion} 

In this paper, we investigated the global structure of the $SL(2,\mathbb{Z})$-invariant potentials describing inflationary $\alpha$-attractors
\cite{Kallosh:2024ymt,KalLin2024,CKLR}.
 These potentials contain an inflationary plateau covering the upper part of the Poincar\'e half-plane. Meanwhile, the lower part of the half-plane is covered by infinitely many sharp ridges, forming a complicated fractal landscape close to the boundary of the half-plane at $\tau_{2}= 0$. However, we found that each sharp ridge in the lower part of the half-plane is physically equivalent to the inflationary plateau in the upper part of the half-plane.

To describe the modular landscape and remove the distortion caused by the factor ${1\over y^2}$ in the half-plane metric, we switched from axion-inflaton coordinates with a kinetic term 
\begin{equation}\label{kin1}
{\cal L}_{kin}^{axion-inflaton}=\frac12(\partial\vp)^2+\frac{3\alpha}{4}e^{-2\sqrt{\frac{2}{3\alpha}}\vp}(\partial\theta)^2 
\end{equation}
to Killing coordinates with the kinetic term 
\be
{\cal L}_{kin}^{Killing}={1\over 2}{ (\partial \tilde\vp)^2+ (\partial\vt)^2\over \cos^2 (\sqrt{2\over 3\alpha} \vt)} \ .
\label{Lkin}\ee
 This coordinate system was very useful for the investigation of $\alpha$-attractors in   \cite{Carrasco:2015rva,Carrasco:2015pla}.  In this paper we used this coordinate system, as well as its generalizations.

 In Killing coordinates, the metric does not depend on the inflaton field $\tilde \vp$, and away from the boundary $\sqrt{2\over 3\alpha} \vt = \pm  
{\pi\over 2}$ both $\tilde \vp$ and $\vt$ are close to canonical variables. 
 Moreover, the 
$SL(2,\mathbb{Z})$ invariant potentials are symmetric under inversion, which implies a simultaneous change of sign of $\tilde \vp$ and $\vt$:
\be
\Phi \to -\Phi\, , \qquad \Phi = \sqrt{2\over 3\alpha} (\tilde \vp -i \vt) \ .
\ee
This symmetry makes the plateaus at negative  $\tilde \vp, \vt$ look like exact copies of the ones at positive $\tilde \vp, \vt$. 

In technical terms, we have first identified the formula defining the proliferation of the saddle points in the landscape  
\be
\tau_n^{saddle} =  {i  \over  c^2 +d^2}+  { bd+ac \over  c^2 +d^2} \ .
\label{saddlen1}\ee

We then performed a specific restricted $SL(2,\mathbb{R})$ transformation, and this gave us a set of coordinate systems that can place {\it any saddle point at the center of the corresponding coordinate system}.
\be
\tau^{\pm} [a,b,c,d]= i {e^{^{\pm} \Phi}\over c^2+d^2} + { bd+ac \over  c^2 +d^2} \ .
\label{CS}\ee
Indeed,  Eq. \rf{CS} shows that at $\Phi=0$  we have a saddle point defined by Eq. \rf{saddlen1}.

 Our results show that every pyramid or shark fin or ridge in Figs. \ref{Cartesian}, \ref{3band}, \ref{5band} is actually a plateau when viewed in a proper Killing coordinate system. For example, our Figs. \ref{RenPot} and \ref{saddles1}  at the center have a  saddle point at $\tau=i$. One can clearly see two plateaus descending to this saddle point. 
 
 Our second example is the saddle point between two minima; one can see it in Fig. \ref{5band}.
 One minimum is at the exit from a plateau; the other is at the exit from a ridge. The same saddle point in Killing coordinates is shown in Fig. \ref{17} at the point $\tilde \vp=\vt=0$. It is a bridge between two minima, and there are two plateaus
descending to these two minima.

The same situation was demonstrated in other examples, including the ridges and saddle points at large negative values of $\vp$, far away from the upper part of the potential, see Figs.  \ref{k2},  \ref{long}. In all of these cases, just as in the simplest case shown in Fig. \ref{RenPot}, the upper plateau extends to infinitely large positive values of $\tilde \vp$. It corresponds to the entire upper part of the original potential at $\vp > 0$, $-\infty < \theta < \infty$, not to the fundamental domain.  Similarly, the lower plateau in all of these figures extends to infinitely large negative values of $\tilde \vp.$ It is physically equivalent to the entire upper plateau, and, therefore, also to the entire upper part of the original potential at $\vp>0$, $-\infty<\theta<\infty$.

This means, in particular, that inflation is possible at each of these plateaus, and inflationary predictions in each of these cases should coincide with the standard predictions of cosmological $\alpha$-attractors \cite{Kallosh:2013yoa,Galante:2014ifa,Kallosh:2021mnu}.  

In conclusion, we should say that the investigation of $SL(2,\mathbb{Z})$-invariant potentials describing cosmological attractors is still a work in progress. For example, Ref. \cite{Casas:2024jbw} describes a single inflationary plateau at $\vp > 0$. 
Then in \cite{Kallosh:2024ymt,KalLin2024}, we found a large family of $SL(2,\mathbb{Z})$-invariant plateau potentials, and found that for each inflationary plateau at $\vp > 0$, there was also an equally good inflationary plateau at $\vp < 0$, as in T-model versions of $\alpha$-attractors. In this paper, we have found that such models describe not just one or two plateaus but infinitely many inflationary plateaus masquerading as sharp ridges. While our understanding of the situation develops so fast, there is always a risk of missing something out, but also a possibility of further interesting generalizations that one may want to explore.

\section*{Acknowledgement}
We are grateful to  J.J. Carrasco,  D. Roest, T. Wrase and Y. Yamada for very useful discussions.
This work is supported by SITP and by the US National Science Foundation Grant   PHY-2310429.

\appendix
\section{Geodesics in Poincar\'e half-plane}\label{Geodesics}

``Poincar\'e half-plane'' is a specific model of hyperbolic geometry where the space is represented by the upper half of an Euclidean plane, with a special metric (the Poincar\'e metric) that distorts distances to model hyperbolic geometry properties; it's a half-plane with a non-Euclidean geometry.

In hyperbolic space and in other Riemannian manifolds, a geodesic is a distance-minimizing path from one point to another. Poincar\'e {\it upper half-plane} is the upper half-plane model with $z=x+iy $ defined as 
\be
{\cal H}= \{ z \in \mathbb{C}  | \, y >0 \}
\ee
The real line $y=0$ is excluded since the metric is $ds^2= {dx^2+dy^2\over y^2}$.
The Poincare half-plane model, which is used to study hyperbolic geometry, is considered geodesically complete, even though its geodesics appear as curves that approach the boundary of the half-plane.

To prove the geodesic completeness of the Poincare half-plane, one can show that for any two points in the half-plane, there exists a geodesic connecting them that can be extended indefinitely, which is the key characteristic of geodesic completeness; this is typically done by explicitly calculating the geodesics as either vertical lines or semicircles centered on the x-axis and demonstrating that they can be extended infinitely in both directions. 
\begin{figure}[H]
\centering
\includegraphics[scale=0.40]{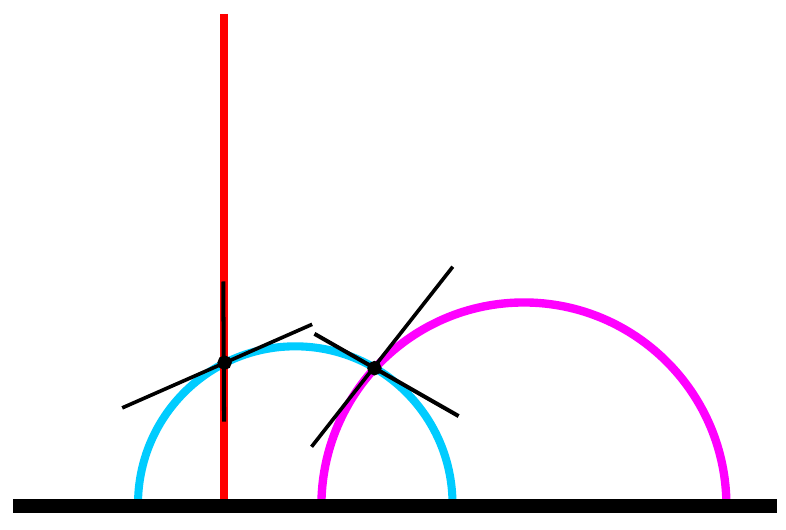}
\vskip -10pt
\caption{\footnotesize The red line is an example of the vertical geodesic. Other examples, blue and purple ones,  are examples of the semicircle geodesics, see https://www.math.brown.edu/reschwar/INF/handout10.pdf }
\label{Geo}
\end{figure}

\noindent {\it Geodesics}: A geodesic in half-plane is either a semicircle meeting the real axis at a right angle or a vertical ray emanating from a point on the real axis. 

\noindent {\it Angle}: The angle between two geodesics is defined to be the angle between the tangents to the geodesics at their intersection point. Fig. \ref{Geo} shows two examples.

\noindent {\it Map} $f(z) = -1/z$ is a symmetry of the hyperbolic plane. The map $f(z) = -1/z$ maps hyperbolic geodesics to hyperbolic
geodesics

\noindent {\it Mobius} \, In the upper half-plane, Mobius transformations map the vertical line through the origin to other vertical lines and to half-circles orthogonal to the real line.

\bibliographystyle{JHEP}
\bibliography{lindekalloshrefs}
\end{document}